\documentclass[a4paper, dvipsnames, 12pt]{article}
\usepackage[warn]{mathtext}
\usepackage[utf8]{inputenc}
\usepackage{tikz}
\usepackage{graphicx}
\usepackage{amsmath,amsthm,amssymb}
\usepackage{alltt}
\usepackage{gensymb}
\usepackage[english]{babel}
\usepackage[top=1.5cm, left=2 cm]{geometry}
\usepackage{wrapfig}
\usepackage{float}
\usepackage{hyperref}
\usepackage{physics}
\usepackage{cancel}
\usepackage[english]{babel}
\usepackage{amsmath,amssymb,amsthm,mathtools,bm}
\usepackage{euscript,mathrsfs}
\usepackage{icomma}
\usepackage{graphicx}
\usepackage{wrapfig}

\usepackage{indentfirst}
\usepackage{amsbsy}
\usepackage{wasysym}
\usepackage{hyperref}
\usepackage{cancel}
\usepackage{verbatim}
\usepackage{empheq}
\usepackage{adjustbox}
\usepackage{authblk}
\usepackage{tikz-cd}

\begin{document}

\title{$\theta$-term in Russian Doll Model: phase structure, quantum metric
and BPS fractality }
\author{Alexander Gorsky$^{1,2}$ and Ilya Liubimov$^{3,4}$}

\affil{$^1$Institute for Information Transmission Problems RAS, 127051 Moscow, Russia \\ 
$^2$Laboratory of Complex Networks, Center for Neurophysics and Neuromorphic Technologies\\
$^3$
Landau Phystech School of Physics and Research, Moscow Institute for Physics and Technology, Dolgoprudnyi,Russia\\
$^4$
Phystech School of Applied Mathematics and Computer Science, Moscow Institute for Physics and Technology, Dolgoprudnyi,Russia
}

\maketitle
\begin{center}
 {\bf Abstract}   
\end{center}

 We investigate the phase structure of the deterministic and disordered versions of the Russian Doll Model (RDM), which is a generalization of Richardson model of superconductivity in a finite system with time-reversal symmetry(TRS) breaking parameter $\theta$. It is one of the simplest examples of the cyclic RG. The deterministic model is integrable and shares the same Bethe Ansatz (BA) equations with the inhomogeneous twisted XXX spin chain.
 Using the BA equation in the single Cooper pair  sector, we analyze the quantum metric, the Berry curvature, and the fractal dimension of RDM eigenstates. A rich structure is found in the parameter plane $(\theta,\gamma)$, where $\gamma \log N$ quantifies the hopping term. 
 For the deterministic RDM we identify the extended domain of the non-ergodic fractal phase on the $(\theta,\gamma)$ parameter plane where the quantum number $Q(\theta,\gamma)$, which arises from the BA equation, exhibits the staircase behavior. The BA equations in RDM exactly coincide with the equations defining the ground states in the theory on the worldvolume of the vortex strings in $N_F=2N_C$ ${\cal N}=2$ SQCD at a strong coupling point $\frac{1}{g_{YM}^2}=0$ with identification $\theta_{RDM}= \theta_{4D}-\pi$. The Hamiltonian of the RDM model is identified with one of the commuting families of non-local
 Hamiltonians of the twisted inhomogeneous XXX spin chain describing the vortex string; hence the exact fractality of the eigenmode of $H_{RDM}$ implies the fractality   in the peculiar 2d-4d BPS sector of the SQCD Hilbert space. Our findings provide an example of the BPS fractality regime for the probe operator in the sector of Hilbert space with some amount of SUSY which is determined by the $N$-scaling of the effective Planck constant.

\newpage
\section{Introduction}

The Russian Doll Model (RDM) has been introduced
in the context of superconductivity in the finite-dimensional system \cite{ Leclair2004russian,leclair2004log, Leclair2003russian,  Dunning2004integrability} 
and is the perfect laboratory to investigate the effects of the time-reversal symmetry (TRS) breaking in finite-dimensional systems. It is a generalization of the Richardson model \cite{Richardson1963restricted, Richardson1964exact,ortiz2005bcs, dukelsky2004colloquium} with unbroken time-reversal symmetry. Both Richardson and RDM are Bethe Anzatz(BA) integrable; 
the BA equations in the Richardson model coincide with those for the twisted $SL(2)$ Gaudin model, while the BA equation in RDM coincide with those for the twisted non-homogeneous XXX $SL(2,R)$ spin chain. The RDM enjoys some interesting properties; it is one of the simplest examples with the cyclic renormalization group
\cite{glazek2002limit} formulated in the RDM in 
\cite{Leclair2004russian} (see \cite{Bulycheva2014spectrum, braaten2006universality} for reviews) which supports the tower of gaps with the Efimov scaling. The global charge $Q$ that defines the tower of localization scales $\Delta_Q$ can be recognized in both the mean-field solution and the BA equations \cite{anfossi2005elementary}. 

The features of the disordered Richardson model were discussed in \cite{buccheri2011structure}. The Richardson model version emerges
if the complex SYK model \cite{gu2020notes} is perturbed by the attractive Hubbard interaction \cite{wang2020sachdev}.
and the superconducting phase arises in the limit of dominance of the Hubbard term. Another fermionic version of the SYK model 
is also reduced to the Richardson model \cite{iyoda2018effective}.
The disordered version of the RDM model has been introduced in \cite{motamarri2022localization} and the single Cooper pair spectrum manifests the
Anderson localization transition and the multifractal non-ergodic extended (NEE) phase in some range of parameters. This phase was first observed in \cite{kravtsov2015random} in the particular disordered model with matrix Hamiltonian.  The cyclic RG for the disordered case is refined \cite{motamarri2024refined} and the period of the cyclic RG becomes energy dependent.

The phase structure in models involving the TRS breaking parameter and Anderson localization has been discussed in \cite{altland2015topology,zhang2023anatomy, cai2013topological} and it was argued that in disordered systems with topological terms there is an interesting substructure in the localized phase, the mean value of the winding number distinguishes the fine structure.
On the other hand, the intermediate multifractal phase has been discussed in deterministic 1d models with quasiperiodic potentials that mimic the disorder.  
The multifractal phase has also been found in the deterministic hopping problem on weighted graphs and interpreted in terms of the Liouville theory \cite{comtet2001multifractality}.
Another example of the multifractal phase in the deterministic model has recently been found in \cite{das2025emergent} in the Hamiltonians in the banded matrix.
Moreover, there are multiple reentrance transitions between the fractal and localized phases in the 1d quasiperiodic potentials \cite{padhan2022emergence, roy2021reentrant}. The number of such reentrance transitions depends on the coupling constants.

The seemingly unrelated research area concerns the investigation of BH microstates at finite $N$ \cite{chang2023words,chang2024decoding,choi2023towards,choi2024finite} and their role
in a possible instability of BH.
 In particular , two different types of BPS states can be classified according to their large N behavior; monotonous and fortuitous. The monotonous states are smoothly stabilized at $N\rightarrow \infty$, while the fortuitous states are related to the breakdown of BPS-ness at finite N. The monotonous states at large N correspond to gravitons and can not form the horizon, while the fortuitous states are candidates for the BH horizon formation. The fortuitous states exhibit the phenomenon when the R-charge is concentrated at some interval or region at finite N.
The classification can be made rigorous using the BPS- cohomology correspondence. The existence of fortuitous states is related to a long exact sequence
\cite{chang2024fortuity}, and the observed phenomena of R-charge concentration and fortuity have been interpreted in the cohomological language \cite{chang2024fortuity}.

Examples of fortuitous states and R-charge restriction or concentration were found in the supersymmetric ${\cal N}=2$ SYK model \cite{fu2017supersymmetric,turiaci2023mathcal,stanford2020jt,heydeman2023phases} that are considered candidates for the description of the $AdS_2$ near-horizon region of BH. If the spectral asymmetry is added to the ${\cal N}=2$ SYK model,
the phase structure is more complicated, and it was argued that at some critical spectral asymmetry the conformal phase becomes unstable and the gap has developed, which means the effective instability of the BH.
The clear example of the model supporting both monotonous and fortuitous states is presented in \cite{heydeman2023phases}. Interestingly, explicit examples of fortuitous $\frac{1}{16}$ states
in ${\cal N}=4$ SYM were found in \cite{budzik2023following}.

It was suggested in \cite{chen2025bps} that investigating the chaotic properties of BPS states at finite $N$ one can make some claims concerning the presence or absence of horizon in the gravity dual.  It is not a simple problem to investigate the chaotic aspect of the BPS sector due to degeneracy, and it was suggested to consider
the statistical properties of the probe operator instead. A type of operator
projected into the particular BPS sector has been suggested in \cite{lin2023holography}, while a more general analysis has been carried out in \cite{chen2025bps}. It was argued that generically 1/16 BPS states are proper candidates for the fortuitous states and, therefore, could serve as the horizon microstates. The change in statistical properties in the BPS sector was attributed to the "invasion" of non-BPS states at some domain of the parameter space.

In this study, we argue that these two research areas are closely related and the RDM model provides an interesting example, which allows us to investigate some aspects of phenomena quantitatively using BA integrability. The paper has two parts; in the first sections
we investigate the phase structure of RDM combining the analysis of the geometry of the two-dimensional parameter space in RDM and BA equations for clean and disordered cases. Having established the phase structure of RDM, in the second part of the study, we formulate our conjecture that RDM describes the particular BPS sector of ${\cal N}=2$ SQCD near the self-dual strong coupling point. The fractality found in RDM is mapped into the BPS fractality in the sector with one vortex string after proper identification of parameters.

One of the tools we will use is the geometric quantum tensor introduced in \cite{Provost_1980},
which quantifies the response of the system to perturbations. It yields the
induced metric on the parameter space as the real part and the Berry curvature
as the imaginary part. It was argued in \cite{Berry_1984} that the singularity of the Berry curvature emerges from the level crossing both in finite-dimensional systems and in field theory. On the other hand, the singularity in the induced metric or, more generally, in the Ricci curvature indicates the quantum phase transition \cite{Zanardi_2006,Zanardi_2007,Kolodrubetz_2013, Kolodrubetz2017,  Alvarez-Jimenez2017}. Different aspects of geodesic flows on the parameter space towards the singular points have been discussed in \cite{Tomka2016,dymarsky}. Moreover, it was conjectured that the integrability of the model under consideration is enough to get the singularity in the quantum metric \cite{kim2023integrability}. 

The matrix Hamiltonians provide the simplest playground for analyzing the geometry of the parameter space by conventional means of matrix models. The origin of a two-dimensional parameter space for the matrix model with chaotic perturbations has been discussed in \cite{Berry_2020} using the matrix model technique
and in \cite{penner2021hilbert} using the SUSY $\sigma$-model representation. The geometry of the two-dimensional
parameter space of the integrable system of random energy type \cite{Derrida_1980} perturbed by two chaotic perturbations has been
analyzed in \cite{sharipov2024hilbert}. It was shown that the integrable point in the two-dimensional parameter
space corresponds to the conical singularity when embedded in 3d contrary to the chaotic unperturbed
Hamiltonian when the embedding surface is smooth at the origin. Moreover, the components of the metric tensor feel the transitions between the localized and delocalized phases of the model. The identification of phases through geodesic lengths in this model was developed in \cite{gill2025geometry}.
In another matrix model, the perturbed Rosenzweig-Porter model investigated in \cite{skvortsov2022sensitivity} metric feels the whole phase diagram
as well.

We will generalize the analysis of the matrix Hamiltonian with two parameters in \cite{sharipov2024hilbert} for RDM, which brings new essential
features to the geometry of the parameter space.
In particular, we found the deformation of the quantum metric by the TRS breaking parameter and the appearance of the
Berry curvature in the two-dimensional parameter space. It is demonstrated that the
conical singularity is closely related with the level crossing in some domain
of the parameter space, and we investigate the dependence on the number of crossed levels. It is shown that the
non-diagonal component of the metric indicates that the embedding cone is deformed.

The metric indicates the presence of several phases in the plane of the $(\theta,\gamma)$ parameters, but does not allow us to properly identify these phases. To this aim, we evaluated the fractal dimension of the single Cooper pair excitation as a function of parameters. It clearly indicates the presence of three phases: localized, multifractal, and delocalized. It was a bit puzzling for a while how the fractality emerges in the deterministic model with BA integrability. Our study provides a possible pattern for this phenomenon when the global charge $Q$ plays a key role. It turns out that in the multifractal phase there are subdomains in the $(\theta,\gamma)$ parameter
plane when only the fixed values of Q defined in the BA equations due to the multivaluedness of the logarithms are available. It is a kind of marginal stability curve phenomenon.

Upon identifying the phase structure of the RDM model, we use the emerging rich picture to formulate the phenomenon of BPS fractality
in the sector of SQCD Hilbert space with some amount of SUSY. Namely, we shall focus on the vortex string or surface operator 
in $\Omega$-deformed $\cal{N}=2$ SQCD with $N_F=2N_C$. The important step in the formulation of BPS fractality is that  the equation for the ground state in the ${\cal N}=(2,2)$ $\sigma$ model on the worldvolume of the vortex string in the $\Omega$ deformed SYM theory in the Nekrasov-Shatashvili limit coincides with the BA equation for the  twisted inhomogeneous XXX spin chain upon proper identification of parameters \cite{nekrasov2009supersymmetric,nekrasov2010quantization}.
The effective twisted superpotential coincides with the Yang-Yang function for the spin chain. The 2d theory on the vortex string on the Higgs branch is also dual to the 4d theory without the vortex string at the particular quantization locus on the Coulomb branch \cite{dorey2011quantization}.

 In the SQCD without the $\Omega$-deformation  vortex strings  solutions at the Higgs branch were  found in \cite{hanany2003vortices,auzzi2003nonabelian}, see \cite{shifman2009supersymmetric} for the review. 
 These strings support the 1/4 BPS states - monopoles in the Higgs phase \cite{tong2004monopoles,shifman2004non}. Recently, a more complicated BPS configuration was found, which is identified as the closed vortex string with four monopoles \cite{ievlev2020string,ievlev2021critical}. Such baryonic states are supported at the strong-coupling self-dual point $\theta_{4d}=\pi, g_{YM}^2=\infty$ in $N_f=4$ $SU(2)$ theory. They are massless exactly at the self-dual point and can condense.

The $H_{RDM}$ has been identified as the first nontrivial non-local Hamiltonian of an inhomogeneous XXX chain from the commuting family 
derived by expansion of the transfer matrix \cite{bork2015particle}. This explains why the BA equations for the RDM model with N levels and for
the inhomogeneous twisted spin chain with N sites are identical after identification of the parameters \cite{Dunning2004integrability}. 
The $\theta$- parameter in RDM is related to the $(\theta_{4d}-\pi)$- parameter in 4d SQCD 
which can be clarified from the analysis of worldsheet theory  of vortex string without $\Omega$ deformation \cite{ievlev2020string, gerchkovitz2018vortex}. The Richardson limit of RDM is related  to SQCD with $\theta_{4d} \rightarrow \pi$. The number of Bethe roots  according Bethe/gauge correspondence  corresponds to the number 
of vortex strings in 4d SQCD \cite{nekrasov2009supersymmetric} hence  fractality found for the one root solution corresponds to the fractal properties of the
single vortex string or surface defect if the tension of the string tends to infinity.
 Importantly, from the 4d point of view, the parameter $\omega$ involved in the $\Omega$-deformation is proportional to $r\sin (\pi -\theta_{4d})$ 
 where the parameter $r$ corresponds to the chemical potential in the 4d theory. In the spin chain, it plays the role of the 
 effective Planck constant. The transition between the ergodic, fractal and localized phases 
 is governed by the parameter $\gamma$ or the N-scaling of the effective Planck constant $\hbar\propto \omega \propto N^{-\gamma}$

We shall focus on the properties of the Hilbert space subsector of a single vortex string in ${\cal N}=2$ SQCD. In general,
the manifold of BPS states with different amounts of SUSY is quite complicated, involving multiple intersecting CMS that can be described in the framework
of spectral networks \cite{gaiotto2013spectral,gadde2014walls,Gaiotto_2012,galakhov2015spectral}. However, at the particular locus of the  moduli space
with $\frac{1}{g^2}=0$, their description is simplified since
the high level of their collinearity takes place and the notion of the BPS graph has been introduced \cite{longhi2018wall,gabella2017bps}. In general,
at these points the 2d and 4d BPS states are collinear, while the combined 2d-4d states form a kind of halo. 
Therefore, using the relation between two models, upon identification of parameters,
we formulate the BPS fractality phenomena in the very specific 2d-4d subsector of SQCD; however, we expect that more fractal BPS states 
with some amount of SUSY in the Hilbert space 
with complicated spectral networks can be found.

The paper is organized as follows. In Section 2 we recall the definition and key properties of the RDM model.
 In Section 3 we consider numerically and analytically the deterministic and disordered versions and identify extended domains of ergodic, multifractal, and localized phases. We will focus on the BA equation for a single Cooper pair and investigate their properties. The important role of the global charge $Q$ is clarified. 
In Section 4 we consider the quantum metric  for the different patterns of disorder. Using the Bethe-Anzatz equation, we obtain asymptotic behavior for a quantum metric.
The  relation between the RDM model and the BPS subsector in worldvolume theory on the vortex string
of ${\cal N}=2$ SQCD at the strong coupling point is discussed in Section 5 and the conjecture of the BPS fractality phenomenon is formulated.
In Discussion we attempt to place our findings in a more generic context and formulate the possible directions of the further research.
The results of the study are summarized in the Conclusion. In the Appendixes, we comment on  technical aspects of our calculations and 
explain the relation of RDM Hamiltonian and  twisted inhomogeneous XXX quantum spin chains.

\section{Overview of RDM Model}
In this section, we recall the definition and key properties of the RDM.
The RDM is a modification of Richardson's Hamiltonian of reduced BCS superconductivity with TRS broken. The Hamiltonian is defined as:
\begin{equation}
    H = \frac{1}{2}\sum_{n, \sigma}(\varepsilon_{n} - x) c_{n\sigma}^{\dagger}c_{n\sigma} - \sum_{n \neq m }r e^{i\theta sign(n-m)} b_n^{\dagger}b_m 
\end{equation}
where $c_{n \pm}^{\dagger},c_{n \pm}$ are creation/annihilation operators of the fermions in time-reversal states $\pm$, $re^{i\theta} = x+iy$, while 
$b_n^{\dagger} = c_{n+}^{\dagger}c_{n-}^{\dagger}$, $b_n = c_{n-}c_{n+}$ are creation/annihilation operators of the Cooper pairs. 
\begin{equation}\label{Hamiltonian}
    H = \sum_{n = 1}^{N}(\varepsilon_n - x)b_n^{\dagger}b_n - \sum_{n \neq m }r e^{i\theta sign(n-m)} b_n^{\dagger}b_m 
\end{equation}
The operators $(n_k= b_k^{\dagger}b_k - 1/2, b_k^{\dagger},b_k)$ form the pseudospin algebra $SU(2)$, supported by the relation for hard-core bosons. 
\begin{equation}
   (b_j^+)^2 =0,\qquad [b_i,b_j]=[b_i^+,b_j^+]=0,\qquad   [b_i,b^{\dagger}_k]=\delta_{ik}(\frac{1}{2}- b_i^{\dagger}b_i)
\end{equation}
Hence, the Richardson model at $\theta \rightarrow 0$ can be considered as a fully connected XX model in the external magnetic field, while for RDM
the second hopping term
involves the TRS breaking parameter $\theta \in S^1$. 
Note that 
\begin{equation}
    \frac{\partial H}{\partial \theta}_{\theta=0} \propto r\sum_k n_k
\end{equation}
We will use another parameterization of $r$ in terms of $\gamma$ and $N$, which will be convenient for determining the phases: 
$$r= N^{-\gamma}$$
which differs by factor 2 from the definition of $\gamma$ in \cite{motamarri2022localization,motamarri2024refined}. 
The model is integrable via the quantum inverse scattering method \cite{Dunning2004integrability} and the
Hamiltonian and other conserved higher Hamiltonians can be derived by expansion of the transfer matrix
of the inhomogeneous twisted XXX spin chain.
Consider the anzatz for the wave functions 
\begin{equation}
    \ket{M} = \prod^{M}_{i = 1} B_i(E_i)\ket{vac}
\end{equation}
where rapidities obey the BA equations. For a model with $N$ sites in the sector with $M$ pairs, the
system of Bethe equations reads 
\begin{equation}
    e^{-2i\theta}\prod_{l = 1}^{N} \frac{E_a - \varepsilon_l - iy}{E_a- \varepsilon_l + iy} = \prod_{b=1}^M \frac{E_a-E_b - 2iy}{{E_a- E_b + 2iy}}
\end{equation}
 taking $\log$ of both parts and choosing branch of the multivalued function, we obtain:
\begin{equation}\label{BAeq}
     \sum_{l = 1}^N \arctan{\frac{y}{E_a - \varepsilon_l}} = \sum_{b\neq a }{\arctan\frac{2y}{E_a - E_b}}-\theta + \pi Q_a ,
\end{equation}
the total energy of the state $E = \sum_{a = 1} E_a $. 
The BA equations have the generating Yang-Yang function ${\cal W}$ :
\begin{equation}\label{super}
    \frac {\partial {\cal W}(E_i,\theta,\varepsilon_i,\gamma)}{\partial E_i}=Q_i
\end{equation}
The same Yang-Yang function for the non-homogeneous twisted XXX chain emerged in the theory of a vortex string in $\Omega$ deformed ${\cal N}$ = 2 SUSY QCD
as the effective twisted superpotential and (\ref{super}) defines the ground states in the 2D sigma model
\cite{nekrasov2009supersymmetric, nekrasov2010quantization}. 
In the limit $y \to 0$ the equation \ref{BAeq} turns into the Richardson equation \cite{anfossi2005elementary} with the effective coupling constant $\frac{1}{G_Q} = \frac{1}{y}(\arctan{\frac{y}{x}}+\pi Q)$:
\begin{equation}\label{effRich}
    \frac{1}{G_Q} + \sum_{l = 1}^N {\frac{1}{E_a - \varepsilon_l}} - \sum_{b\neq a }{\frac{2}{E_a - E_b}} = 0
\end{equation} 
The Hamiltonian of the Richardson model can be expressed in terms of the conserved commuting Hamiltonians of the Gaudin model.
\begin{equation}
    R_i= -t_i -2G\sum_j \frac{t_it_J}{\varepsilon_i -\varepsilon_j}
\end{equation}
as follows 
\begin{equation}
    H_{Rich}=\sum_i \varepsilon_i R_i + G(\sum_i R_I)^2 + const 
    \label{richR}
\end{equation}
Similarly, the RDM Hamiltonian can be derived from the expansion of the transfer matrix
of the inhomogeneous  twisted  XXX spin chain  \cite{Dunning2004integrability, bork2015particle}. We describe the
emergence of the $H_{RDM}$ from the expansion of transfer matrix in the Appendix.

The BAEs for Gaudin and inhomogeneous XXX chains emerge in the semiclassical limits of the KZ equations for the WZW conformal blocks \cite{reshetikhin1994quasiclassical}. In the
context of the Richardson model, the KZ formulation has been developed in \cite{sierra2000conformal} and a recent discussion of this issue
can be found in \cite{biskowski20252dcftefficientbethe}. The link with the peculiar irregular conformal blocks investigated in \cite{gaiotto2012knot}
in the context of knot invariants was discussed there. Very explicit expressions for the wave functions of the inhomogeneous XXX spin chains have been found in \cite{lee2020quantum, nekrasov2022surface} while the brane representation in the context of 4d Chern-Simons theory is developed in \cite{costello2020unification}.

Similarly to the BCS solution, a gap is formed for the model with the interaction of electrons inside the Debye shell: $\abs{\varepsilon_j} < \omega_c $. The gap equation can be obtained via the mean-field approximation or from equation (\ref{effRich}): 
\begin{equation}
     \frac{1}{G_Q} = \frac{1}{y}(\arctan{\frac{y}{x}}+\pi Q) = \int_0^{\omega_c}d\varepsilon \frac{N(\varepsilon)}{\sqrt{\Delta^2+ \varepsilon^2}} 
\end{equation}
Provides a specific RDM scaling of the gaps.
\begin{equation}
\Delta=\Delta_0e^{-Q\lambda}
\end{equation}
where $\lambda= \frac{\pi \delta}{r sin \theta}$ corresponds to the period of the RG. The value of $Q$ changes by one during the single RG cycle. Higher values of $Q$ correspond to larger sizes of Cooper pairs that behave as 
\begin{equation}
    r = r_0 e^ {\lambda Q}
\end{equation}
In what follows, we shall focus on the $M=1$ case corresponding to the single Cooper pair. For the one-pair sector, we have a matrix Hamiltonian:
\begin{equation}
    H_{nm} = \delta_{nm} (\varepsilon_m - x) - r e^{i\theta sign(n-m)}
\end{equation}
with the Bethe equation: 
\begin{equation}\label{BAeq1}
     \sum_{l = 1}^N \arctan{\frac{y}{(E  - \varepsilon_l)}} = -\theta + \pi Q 
\end{equation}

\section{Phase structure of RDM}
\subsection{Fractality for clean RDM model}
The starting point is the exact solution for the eigenstates in the one-pair sector, which is presented in the Appendix. The eigenstates, expressed as functions of $\varepsilon_i, y$ and $E$, are as follows:
\begin{equation}\label{exaxt_main}
    \psi_j =  \frac{1}{\sqrt{\sum_{k} \frac{1}{\rho^2_k}}}\frac{1}{\sqrt{(E -\varepsilon_j)^2 + y^2 }}\exp({-i(\varphi_{j} + \varphi_1+ 2\sum_{k = 2}^{j-1} \varphi_k)})
\end{equation}
with parameters $\rho_i = \sqrt{(E -\varepsilon_i)^2 + y^2 }$, $\varphi_i = \arctan{\frac{y}{E- \varepsilon_i}}$. 
The eigenstates have a Lorentzian structure with $\Gamma = y = r\sin{\theta}$. This form of $\abs{\psi_i}$ coincides with the perturbation theory ansatz, proposed to describe fractal states \cite{monthus2017multifractality}, however, in our case, it is an exact solution. In the limit $\theta\to 0$, which corresponds to the Richardson model, $\Gamma$ vanishes. In case of trivial diagonal elements $\varepsilon_i = 0$, the eigenstates have the form of plane waves with momentum $p = \frac{2i(\pi Q - \theta)}{N}$:
\begin{equation}\label{plain_main}
\ket{Q} = \sum_n \frac{\exp(-\frac{2i(\pi Q - \theta)}{N} n)}{\sqrt{N}} \ket{n} 
\end{equation}
We now demonstrate exactly the presence of fractal phase for the deterministic RDM. Recall the definition of the fractal dimension $D_q$ for an eigenstate $\psi_n(i)$:
\begin{equation}\label{fractal_def}
    I_q = \sum_{i} |\psi_n(i)|^{2q} \sim N^{D_q (1-q)}
\end{equation}
Then, in the large $N$ limit fractal dimension can be evaluated as:
\begin{equation}
    D_q = \frac{1}{1-q}\frac{\ln{I_q}}{\ln{N}}
\end{equation}
We will now use an explicit form of the eigenstates to calculate $I_q$ and demonstrate their fractality, leaving technical details to Appendix, we can find the fractal dimension in the region $\gamma \in (0,1)$:
\begin{equation}\label{D_q_clean}
    D_q = 1 - \gamma + \frac{\ln{\sin{\theta}}}{\ln{N}} + \frac{\ln{\pi}}{\ln{N}} + \frac{1}{2}\frac{1}{q - 1}\frac{\ln{\pi}}{\ln{N}} + \frac{1}{q-1}\frac{\ln{\frac{\Gamma(q)}{\Gamma(q - 1/2)}}}{\ln{N}}
\end{equation}
Thus, the RDM exhibits three distinct phases:
\begin{itemize}
    \item a localized phase for $\gamma > 1$, $D_q = 0$. In this phase diagonal elements of Hamiltonian dominate off-diagonal ones.
    \item a fractal phase for $\gamma \in (0, 1)$, $D_q = 1- \gamma$. In this phase both diagonal and off-diagonal elements are important.
    \item a delocalized phase for $\gamma < 0$, $D_q = 1$. In this phase off-diagonal elements dominate diagonal ones, and eigenstates are close to plain waves (\ref{plain_main}).
\end{itemize}
\subsection{Scaling properties of critical points}
Let us now turn to general disordered case, with $\varepsilon_i$ uniformly distributed in $\in [-W/2+\delta(i-N/2), W/2+\delta(i-N/2)]$. In the previous section, we observed that the two points where the behavior of the deterministic model with $\delta \sim \omega/N$ changed were $\gamma = 0$ and $\gamma = 1$. We will now consider the general values of $\delta$, the width of the level distribution $W$, and the parameter $r$. Since the matrix elements $H_{nm}$ are linear in these parameters, the Hamiltonian can be rescaled $H(W, \delta, r) = \lambda H(W/\lambda, \delta/\lambda, r/\lambda) = \lambda \tilde{H}(\tilde{W}, \tilde{\delta}, \tilde{r})$. The eigenstates and their fractal dimensions are invariant under this transformation. Therefore, for new points we obtain the following relation:
\begin{equation}\label{scaling}
    {\gamma}_{cr}(W/\lambda, \delta/\lambda) = \gamma_{cr}(W,\delta) + \frac{\ln{\lambda}}{\ln{N}}  
\end{equation}
For our numerical calculations in deterministic case, we used $\delta = 1$. This choice implies that the critical point will be shifted by 1 in terms of $\gamma$:
\begin{equation*}
    \gamma_{\rm{fractal\to loc}}(\delta = \frac{1}{N}) = 1 \rightarrow \gamma_{\rm{fractal\to loc}}(\delta = 1) = 0
\end{equation*}
It was shown in \cite{Khaymovich2023fractalset} that critical points depend only on the fractal dimension of the set in which $\varepsilon_i$ is distributed both in deterministic and disordered cases.
\subsection{Quantum number $Q$ and numerics} 

We now compare the analytical result (\ref{D_q_clean}) with the numerical calculations for deterministic RDM.
\begin{figure}[H]
    \centering
    \includegraphics[width=0.8\linewidth]{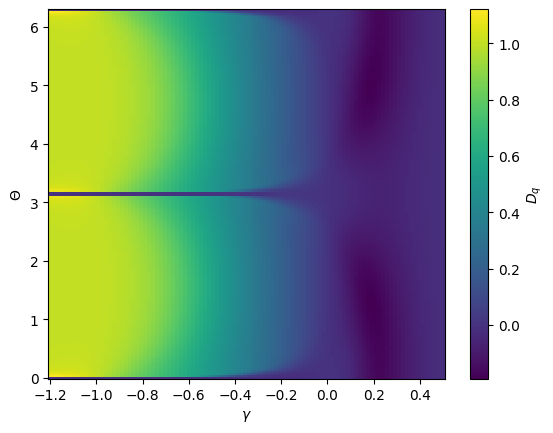}
    \caption{$D_q$ on a ($\gamma, \theta$) plane for $q = 1.75$ for level in the bulk of spectrum} 
    \label{imageD_q}
\end{figure}
To identify the fractal phase, we plot the dependence of $D_q$ on $\gamma, \theta$ in Fig. \ref{imageD_q}. It shows a clear-cut fractal domain with $0<D_{1.75}<1$ in the parameter plane. Although there are minor differences between the bulk and edge states, both exhibit fractality.
\begin{figure}[H]
    \centering
    \includegraphics[width=0.8\linewidth]{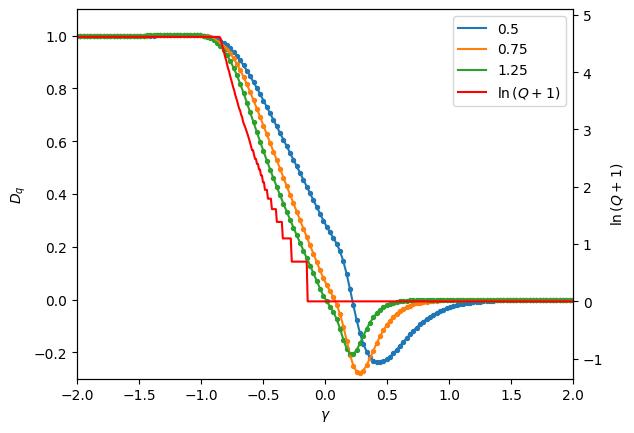}
    \caption{$q$-dependence in $D_q(\gamma)$ for RDM without disorder, averaged for $N = 500 - 1000$ and $\ln{Q}$ staircase structure for level $n = 125$, with $\theta = \pi/4$.} 
    \label{D_for_clean}
\end{figure}
Recall the definition of an integer number $Q$, arising from taking $\ln$ of BAE:
\begin{equation}
    Q  = \frac{\theta}{\pi} + \frac{1}{\pi}\sum_{i=1}^N \arctan{\frac{y}{(E - \varepsilon_i)}}
\end{equation}
 Fig.\ref{D_for_clean} shows that the envelope of the $\ln Q(\gamma)$ plot follows the $D_q(\gamma)$ dependence and can therefore be used as a reliable phase identifier. Moreover, in the deterministic RDM $Q$ is quantized and the graph $Q(\gamma)$ shows the structure of the staircase. We will provide analytical arguments to support this connection. Taking the limit of large $N$ (for $\delta = \omega/N$) we can replace the sum with an integral. This approximation is valid when the characteristic scale of the function is larger than the partition mesh: $y  = \sin{\theta}/N^{\gamma}\gg 1/N$. This condition holds in both the fractal and delocalized phases ($\gamma < 1$):
\begin{equation}\label{Q_int}
    Q  = \frac{\theta}{\pi} + \frac{N}{\pi\omega}\int_{-\omega/2}^{\omega/2} d\xi\arctan{\frac{y}{(E - \xi)}}
\end{equation}
\[
Q =\frac{\theta}{\pi}+ \frac{N}{\pi}\frac{(E + \frac{\omega}{2})}{\omega} \arctan{\frac{y}{E+\frac{\omega}{2}}} - \frac{N}{\pi}\frac{(E - \frac{\omega}{2})}{\omega} \arctan{\frac{y}{E -\frac{\omega}{2}}} - 
\]
\begin{equation}
   -\frac{yN}{2\pi\omega}\ln{\frac{(E -\omega/2)^2+y^2}{(E +\omega/2)^2+y^2}}
\end{equation}
With regimes:
\begin{equation}
    Q = \begin{cases}
        \sim N^{1-\gamma}, \;\;\; \gamma \in (0,1)\\
        \sim N, \;\;\; \gamma < 0 
    \end{cases}
\end{equation}
Let us now plot the $Q$ dependence on the $(\gamma, \theta)$ plane to compare it with Fig. \ref{imageD_q}.
\begin{figure}[H]
    \centering
    \includegraphics[width=0.8\linewidth]{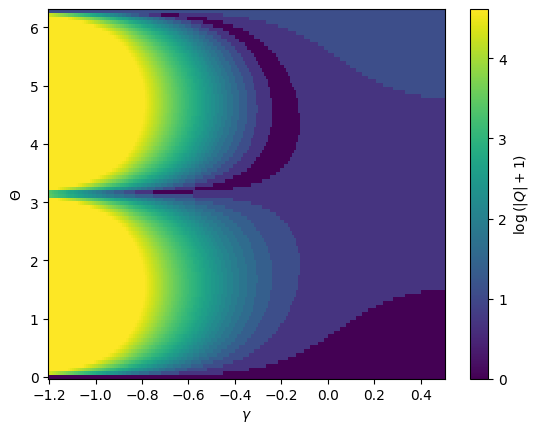}
    \caption{$\ln{(\abs{Q}+1)}$ for a fixed level on a ($\gamma, \theta$) plane } 
    \label{fig:Q_on_plane}
\end{figure}
The fine structure of the plot $Q(\gamma,\theta)$ in Fig. \ref{fig:Q_on_plane} clearly shows the domains with the different values of $Q$ in the $(\gamma,\theta)$ plane. Note that this fine structure is not captured by the $D_q(\gamma,\theta)$ plot. By fixing a level number $n$ (e. g. $n = 125$ in Fig.\ref{D_for_clean} and Fig.\ref{fig:Q_on_plane}, solving the spectral problem numerically and considering the single pair BA equation for $E_n$, we obtain the corresponding $Q$ for every point on the $(\gamma, \theta)$ grid:
\begin{equation}\label{BAeq2}
     \frac{1}{\pi}\sum_{l = 1}^N \arctan{\frac{y}{(E_n (\gamma)- \varepsilon_l)}} +\frac{\theta}{\pi} = Q(n, \gamma, \theta) 
\end{equation}
In the following, we discuss the analytical arguments in support of this conjecture. The yellow regions correspond to the delocalized phase, while the blue regions correspond to the localized phase.  The curves separating different phases can vary for different energy levels, in the limit $N\to \infty$ these differences vanish and the model exhibits localized, fractal and delocalized phases.

It is also important to note that if the energy set contains a solution with $Q_{max} = p$, then there are solutions with all $\abs{Q} < p$ concentrated at the edges and in the middle of the spectrum near $E = 0$. For these levels, the staircase structure is not strongly expressed. Some level-invariant function, such as $\max_{n} Q(E_n)$ can be used to distinguish between different phases.  

On the lines $\theta = 0$ and $\theta = \pi$ the model reduces to the Richardson limit and the fractal phase disappears.
The one-pair case in RDM can be interpreted as the hopping problem in the Fock space for some interacting many-body system
along the framework suggested in \cite{altshuler1997quasiparticle}. It was argued that the one-particle localization in the Fock space indicates the many-body localization in the real space in the interacting system. The Fock space is represented as the full graph in the RDM case which is the simplest example of the regular random graph (RRG) for the degree $d=N-1$. The Anderson localization on the Bethe tree has been analyzed in \cite{abou1973selfconsistent}
and the review of the Anderson localization on the RRG can be found in \cite{tikhonov2021anderson}. In the large $d$ limit, the critical disorder with flat distribution is as follows. 
\begin{equation}
    W_c\propto d\log d
\end{equation}
and there is no fractal phase in the disordered RRG without $\theta$-deformation. As we have observed, the $\theta$-parameter yields a rich phase structure in the RDM.
\subsection{Phase structure in general case}
In the clean RDM, we observed a relation between the behavior of $Q$ and the fractal dimension $D_q$ of the eigenstates. We now consider arguments that support this relation and extend it to the disordered case. The integer charge $Q$ comes from the multivalued function:
 \begin{equation}\label{BAE}
  \pi Q - {\theta}= \sum_{l = 1}^N \arctan{\frac{y}{(E-\varepsilon_l)}}
\end{equation}
\begin{figure}[H]
\begin{minipage}[h]{0.49\linewidth}
\center{\includegraphics[width=1\linewidth]{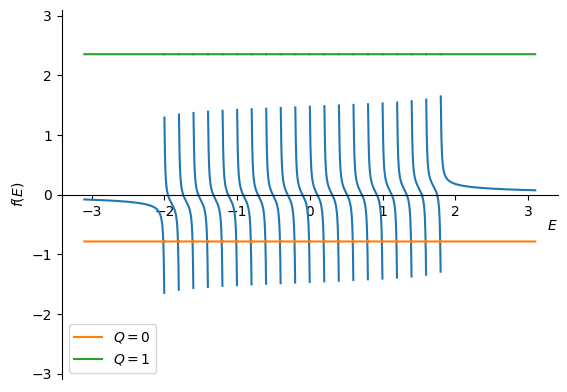} \\a) }
\end{minipage}
\hfill
\begin{minipage}[h]{0.49\linewidth}
\center{\includegraphics[width=1\linewidth]{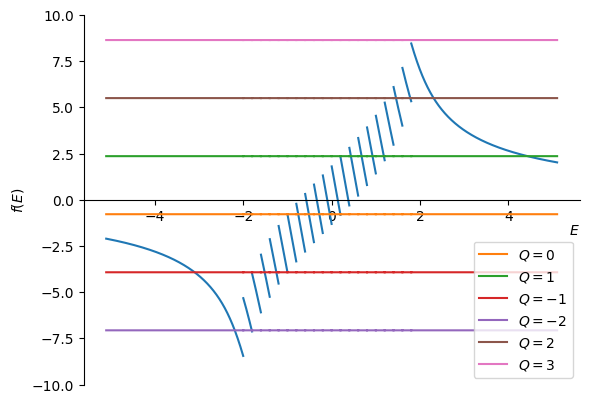} \\b)}
\end{minipage}
\caption{Plot of rhs and lhs of \eqref{BAE} vs $E$ for $N = 20$,  $\delta = 0.4$, discontinuities correspond to $E = \varepsilon_l$ a) localized phase with $\varepsilon_{i+1}-\varepsilon_i = \delta \gg y$ b) fractal phase with $\varepsilon_{i+1}-\varepsilon_i = \delta \ll y \ll N$, where levels with  $Q > 1$ arise}
\label{ris:resonances}
\end{figure}
Each term in the rhs of \eqref{BAE} has discontinuities in $E = \varepsilon_i$, while in the lhs we have a set of constants, indexed by $Q$. Fig. \ref{ris:resonances} shows the typical picture of the level distribution of $Q$ for the deterministic RDM in the localized and fractal phases. In the localized phase, all solutions have the corresponding $Q = 0$, while in the fractal phase, solutions with higher values of $Q$ emerge. To extend this to the disordered case, we treat the energy $E$ as a parameter and average $Q$ over the distribution of diagonal elements.
\begin{equation}\label{meanBAE}
    \langle Q\rangle  = \frac{\theta}{\pi} + \frac{1}{\pi}\sum_{l=1}^N\langle \arctan{\frac{y}{(E- \varepsilon_l)}}\rangle_{\varepsilon}
\end{equation}
Using uniform distribution of $\varepsilon$, for the mean value of $\arctan$ one obtains:
\[
\langle \arctan{\frac{y}{(E- \varepsilon_l)}}\rangle_{\varepsilon} = \frac{(E-\langle\varepsilon_i\rangle+ \frac{W}{2})}{W} \arctan{\frac{y}{E -\langle\varepsilon_i\rangle+\frac{W}{2}}} - 
\]
\begin{equation}\label{mQ}
   - \frac{(E -\langle\varepsilon_i\rangle - \frac{W}{2})}{W} \arctan{\frac{y}{E -\langle\varepsilon_i\rangle-\frac{W}{2}}} - \frac{y}{2W}\ln{\frac{(E-\langle\varepsilon_i\rangle-W/2)^2+y^2}{(E-\langle\varepsilon_i\rangle+W/2)^2+y^2}}
\end{equation}
The fractal dimension reads: 
\begin{equation}
    D_q = d = \frac{\ln{Q}}{\ln{N}}
\end{equation}
Note that $E$ in \eqref{mQ} is a parameter and for different $\gamma$ the characteristic energy scale can be different. In the fractal and localized regimes, we expect the energies to have the same scale as the diagonal elements $E \sim \varepsilon$. As $\gamma$ decreases, the off-diagonal elements begin to dominate over the diagonal ones, and this behavior corresponds to the delocalized phase, where the energy scales with $\gamma$ as given in \eqref{E_free} in Appendix.

\subsection{Strong disorder}
Now consider the case of strong disorder, assuming $\langle \varepsilon_i\rangle = 0$ and $W\sim 1$, therefore, all terms in (\ref{meanBAE}) are the same. Note that in this case $\langle Q \rangle$ coincides with (\ref{Q_int}), after replacing $\omega$ with $W$, which is a consequence of the uniform distribution of diagonal elements. Since $y = r\sin{\theta} = \frac{1}{N^{\gamma}}\sin{\theta}$, we can expand $\arctan$ for small $y$, which corresponds to $\gamma > 0$:
\begin{equation}
    \langle Q\rangle = \frac{\theta}{\pi} + \frac{yN}{2\pi W}\ln{\frac{(E+W/2)^2}{(E-W/2)^2}} + O(y^3)
\end{equation}
Consequently, $\langle Q\rangle$ has the form of $A N^{1-\gamma}\sin{\theta}+B\theta$, and there are two different regimes: for $\gamma > 1$ only a constant survives, hence $\ln{Q}/\ln{N} \to 0$ and $D_q = 0$. For $\gamma \in [0,1]$, $\ln{Q}/\ln{N}$ behaves as $1 - \gamma$. In the limit $\theta \to 0$ ($\theta \ll 1/N^{1 - \gamma}$), that corresponds to Richardson model, $\ln Q/\ln N \to 0$, the fractal phase does not survive. 
\begin{figure}[H]
    \centering
    \includegraphics[width=0.8\linewidth]{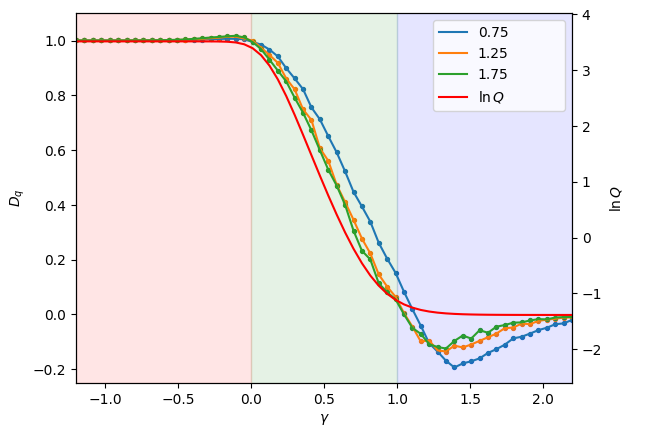}
    \caption{Numerics for $ D_q(\gamma)$ RDM with strong disorder for level $E = 1/3$} 
    \label{fig:my_label1}
\end{figure}
The region of $\gamma < 0$, where $Q \sim N$, can be associated with the delocalized phase.
To complete this section, we consider \eqref{mQ} as a general $W \sim N^p$ and $\langle\varepsilon_i\rangle = 0$. We can use (\ref{scaling}) with $\lambda = W$, $\delta = 0$. We know from the above consideration the fractal-localized transition point for the model with $W\sim 1$:
\begin{equation}
    \gamma(1,0) = p+\gamma = 1
\end{equation}
and delocalized-fractal transition point:
\begin{equation}
    \gamma(1,0) = p+\gamma = 0
\end{equation}
Therefore, for fixed $\gamma$, increasing the diagonal disorder leads first to fractal- delocalization transition at $p_{cr1} = - \gamma$ and then to fractal- localization transition at $p_{cr2} = 1 - \gamma$.

\section{Quantum metric}
We now turn to the quantum geometric tensor, which quantifies the system's response to perturbations. Recall the definition of a quantum geometric tensor in some multidimensional parameter space $\mathbf{\lambda} = {\lambda_1, \lambda_2, ...}$
Consider the eigenvalue equation
\begin{equation}
H(\mathbf{\lambda})\ket{\psi_n(\mathbf{\lambda})} = E(\mathbf{\lambda})\ket{\psi_n(\mathbf{\lambda})} 
\end{equation}
then, the distance between neighbor states is defined as:
\begin{equation}
    ds^2 = 1 - \abs{\bra{\psi_n(\lambda)}\ket{\psi_n(\lambda+d\lambda)}}^2 = g_{\alpha \beta}^{(n)} d\lambda_\alpha d\lambda_\beta
\end{equation}
which reads as:
\begin{equation}\label{g_def}
    g_{\alpha\beta}^{(n)} = (\partial_\alpha\bra{ \psi_n})(\partial_\beta\ket{ \psi_n}) - \partial_\alpha(\bra{ \psi_n})\ket{\psi_n}\bra{ \psi_n}\partial_\beta(\ket{ \psi_n}) = \sum_{m \neq n} \frac{\bra{\psi_n} \partial_\alpha H \ket{\psi_m} \bra{\psi_m} \partial_\beta H\ket{\psi_n}}{(E_n - E_m)^2}
\end{equation}
The real part is a quantum metric tensor, while its imaginary part is the Berry curvature of the energy level $n$. We will use averaged over Hilbert space quantum tensor:
\begin{equation}
    G_{\alpha\beta} = \frac{1}{N}\sum_{n = 1}^N g_{\alpha\beta}^{(n)}
\end{equation}
 Remark that adding a scalar matrix $\Lambda = \mathrm{diag}(\lambda, ... \lambda)$ to the Hamiltonian does not change the metric:
\begin{equation}\label{scalar}
    H \to H + \Lambda \Rightarrow \begin{cases}
        \ket{\psi_n} \to \ket{\psi_n}\\
        E_n \to E_n + \lambda
    \end{cases}
    \Rightarrow G_{\alpha\beta} \to G_{\alpha\beta}
\end{equation}

\subsection{Deterministic case}

For RDM with deterministic diagonal elements: $\varepsilon_n = (n - \frac{N}{2})\delta$, the Hamiltonian can be written as $H = H_0 + V$, where 
\begin{equation}
    H_0 = \sum_{n} \varepsilon_n \ket{n}\bra{n} \;\;\;\;\; V = -\sum_{n \neq m} r e^{i \rm{sign}(n-m)} \ket{n}\bra{m}
\end{equation}
For sufficiently small $r$ we expect the zero-order eigenstates to be localized: $\ket{\psi_n^0} = \ket{n}$, with the corresponding energies: $E^0_n = \varepsilon_n$. Therefore, the quantum metric in the vicinity of $r=0$ can be calculated to leading order using perturbation theory. Leaving technical details to Appendix, we obtain:
\begin{equation}
    G_{rr} = \frac{\pi^2}{3\delta^2}\;\;\;\; G_{\theta\theta} = \frac{\pi^2}{3\delta^2}r^2\;\;\;\;G_{r\theta} = \tilde{A}\frac{\ln^4{N}}{\delta^4}r^5\sin{2\theta}
\end{equation}
Non-zero $G_{r\theta}$ is a manifestation of the time-reversal symmetry breaking.
 \begin{figure}[h]
\begin{minipage}[h]{0.32\linewidth}
\center{\includegraphics[width=1\linewidth]{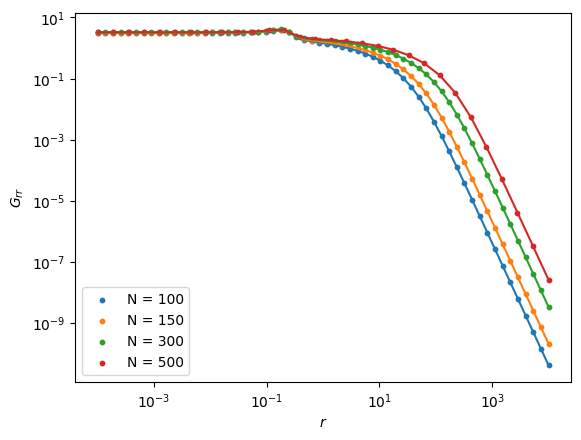} \\ }
\end{minipage}
\hfill
\begin{minipage}[h]{0.32\linewidth}
\center{\includegraphics[width=1\linewidth]{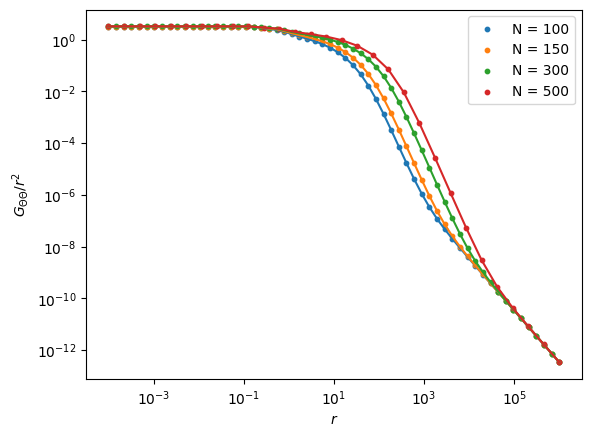} \\}
\end{minipage}
\hfill
\begin{minipage}[h]{0.32\linewidth}
\center{\includegraphics[width=1\linewidth]{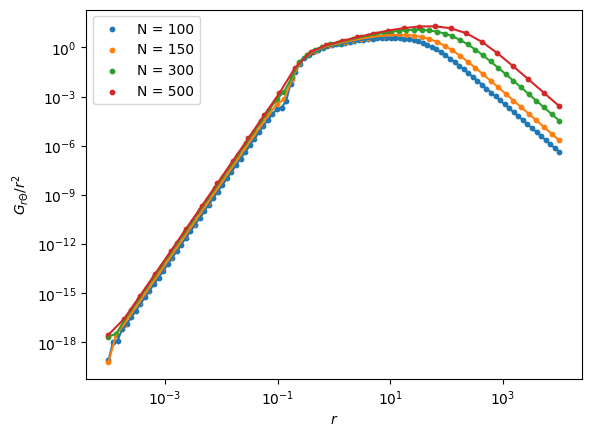} \\}
\end{minipage}
\caption{$G_{\alpha\beta}$ dependence of $r$ for different $N$ for deterministic $\varepsilon_i$}
\label{QGT_det}
\end{figure}
Note that this approximation is valid for $\gamma > 0$, because the terms in perturbation series have the form of $\ln^m{N}$, as seen in $G_{r\theta}$. In parameterization $r = \frac{1}{N^\gamma}$, these terms scale as $\sim \ln^m{N}/N^{n\gamma}$, ensuring the convergence of the series.

\subsection{Weak disorder}

To discuss the case of diagonal disorder for large $N$, we first extract information from Bethe equation (\ref{BAeq1})
assuming that the distribution of diagonal elements is $\varepsilon_i \in [-W/2 + \delta(i-N/2), W/2 + \delta(i-N/2)]$ with $W \in(\delta, 2\delta)$. Recall that in the sum (\ref{BAeq1}) the term is important when $\abs{E - \varepsilon_l}  \sim y$, which means that roots should be localized near $\varepsilon_l$ for small enough $r$. As demonstrated for the $N = 2$ case, see Appendix for technical details, the regime where $E_1 - E_2 \sim r$ occurs when $\varepsilon_{1} - \varepsilon_{2} \sim r$, providing the dominant contribution to the integral and leading to a divergence in the limit $r \to 0$. The denominators $\frac{1}{(E_i-E_j)^2}$ in the QGT definition decompose into two distinct cases: for $\abs{i-j} > 1$ we have $\frac{1}{(E_i-E_j)^2} \leq \frac{1}{\delta^2}$, while for $j = i + 1$ the denominators can be small and lead to a singularity in the limit $r\to 0$. To investigate the analog of singular behavior for large $N$, we consider the assumption $\varepsilon_l - \varepsilon_{l-1} \sim r$. All other terms in BAE can be approximated by their linear parts due to level spacing:
 \begin{equation}
    \theta - \pi Q + \sum_{k = 1}^{l-2} \frac{y}{(E - \varepsilon_k)} + \arctan{\frac{y}{(E - \varepsilon_{l - 1})}} + \arctan{\frac{y}{(E - \varepsilon_l)}} + \sum_{k = l+1}^{N} \frac{y}{(E - \varepsilon_k)} = 0
\end{equation}

  Let $E^{l}_{1,2}$ denote the energy levels of a $N = 2$ RDM with diagonal elements $\varepsilon_1 = \varepsilon_{l-1}$ and $\varepsilon_2 = \varepsilon_{l}$. We expand the roots in the form: $E = E^{l}_{1,2} + \Delta E$ and find the first correction $\Delta E$:

\begin{equation}
   \sum_{k = 1}^{l-2} \frac{y}{(E^{l}_{1,2} - \varepsilon_k)} -\frac{y \Delta E}{(E^{l}_{1,2}- \varepsilon_{l-1})^2 + y^2} -\frac{y \Delta E}{(E^{l}_{1,2}- \varepsilon_{l})^2 + y^2} + \sum_{k = l+1}^{N} \frac{y}{(E^{l}_{1,2} - \varepsilon_k)} = 0
\end{equation}
From the $N = 2$ case, we know that $(E^{l}_{1,2} - \varepsilon_{l-1})$ and $(E^{l}_{1,2} - \varepsilon_{l})$ are of the order $r$, and the same holds for $y$. Therefore, $\Delta E \sim r^2$ confirming the self-consistency of the approximation. The equation for resonant levels reads as follows: 
\begin{equation}
    \theta - \pi Q + \arctan{\frac{y}{(E - \varepsilon_{l - 1})}} + \arctan{\frac{y}{(E- \varepsilon_l)}}  = 0
\end{equation}
and is identical to the result from degenerate perturbation theory. However, the standard degenerate perturbation theory requires $V\gg E_i - E_j$, while in our case $V\sim E_i - E_j \sim r$. The QGT contains two distinct types of terms: $\bra{l-1} ...\ket{l}$ and $\bra{l-p} ...\ket{l}$, $p > 1$. For the latter, the level spacing prevents singular behavior, allowing perturbation theory to be applied directly. To complete our consideration, we need to comment on the eigenvectors: for $E_{l-1}$ and $E_l$ the eigenvectors have the corresponding $2d$ vector components at the $l-1$ and $l$ sites. Then $H\ket{l_{1, 2}} = E^{l}_{1,2}\ket{l_{1, 2}} + O(r)$, and this correction gives a constant contribution to the metric after integration. Using the fact that an additional scalar matrix does not change QGT, we obtain the result that the singular part of the metric comes from $2(N-1)$ terms:
\begin{equation}
    G_{rr} = \frac{\pi}{2rW^2}(W - \delta) 
\end{equation}
\begin{equation}
    G_{\theta\theta} = \frac{\pi r}{W^2}(W - \delta) 
\end{equation}
For $G_{r\theta}$ $N = 2$ the contribution to the metric vanishes, and the result is the same as in the clean model.
\begin{equation}
    G_{r\theta} = \widetilde{A}\frac{\ln^4{N}}{\delta^4}r^5 \sin{2\theta}
\end{equation}

\begin{figure}[h]
\begin{minipage}[h]{0.32\linewidth}
\center{\includegraphics[width=1\linewidth]{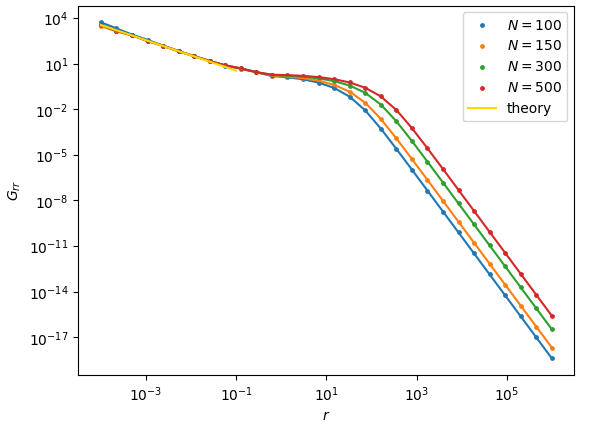} \\ }
\end{minipage}
\hfill
\begin{minipage}[h]{0.32\linewidth}
\center{\includegraphics[width=1\linewidth]{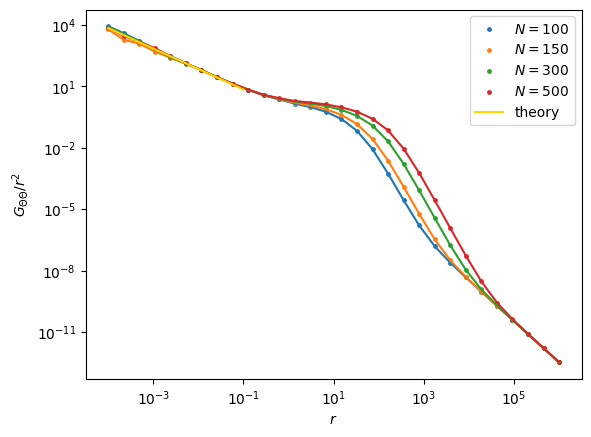} \\}
\end{minipage}
\hfill
\begin{minipage}[h]{0.32\linewidth}
\center{\includegraphics[width=1\linewidth]{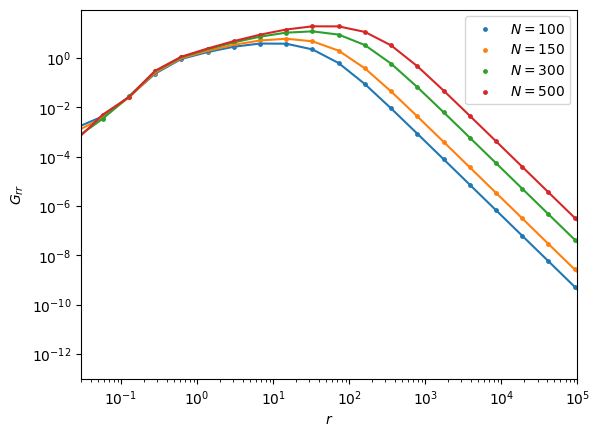} \\}
\end{minipage}
\caption{$G_{\alpha\beta}$ dependence of $r$ for different $N$ in case of weak disorder for $W = 1.5$, $\delta = 1$}
\label{QGT_weak}
\end{figure}
\begin{figure}[H]
    \centering
    \includegraphics[width=0.5\linewidth]{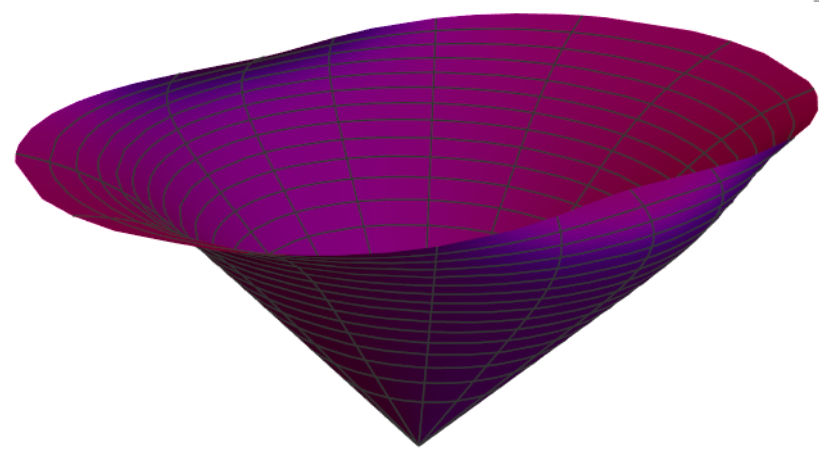}
    \caption{Embedding of isometric manifold for $N \gg 1$ and $\delta = 1$} 
    \label{fig:my_label2}
\end{figure}
We obtain a singular metric at $r = 0$, and the two-level interaction approximation works in a localized phase up to a region in the vicinity of a localized-fractal transition point. The phase structure is the same as in the disorder-free case.

\subsection{Strong disorder}

 \begin{figure}[H]
\begin{minipage}[h]{0.49\linewidth}
\center{\includegraphics[width=1\linewidth]{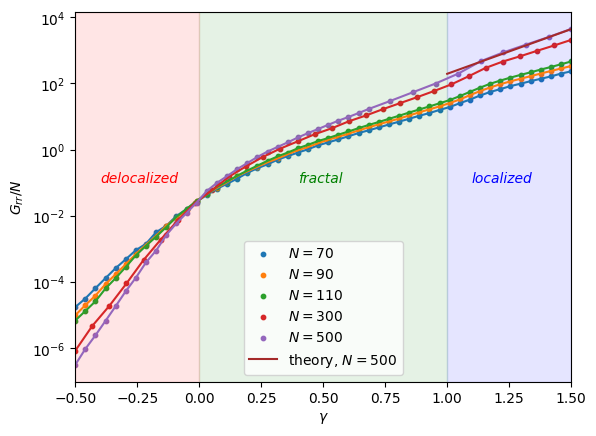} \\ }
\end{minipage}
\hfill
\begin{minipage}[h]{0.49\linewidth}
\center{\includegraphics[width=1\linewidth]{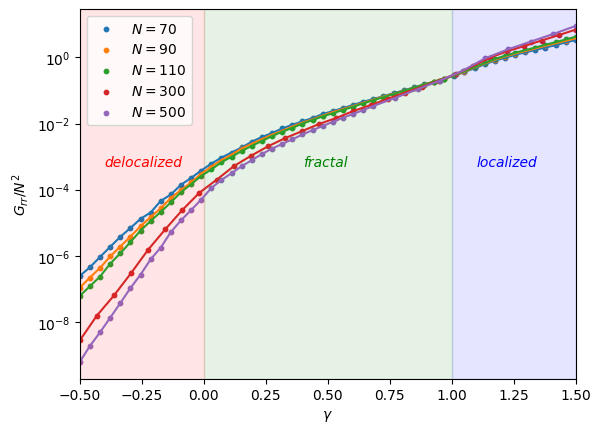} \\}
\end{minipage}
\caption{$G_{rr}$ for $r = 1/N^{\gamma}$ for different $N$}
\label{fig:G_rr}
\end{figure}
 \begin{figure}[H]
\begin{minipage}[h]{0.49\linewidth}
\center{\includegraphics[width=1\linewidth]{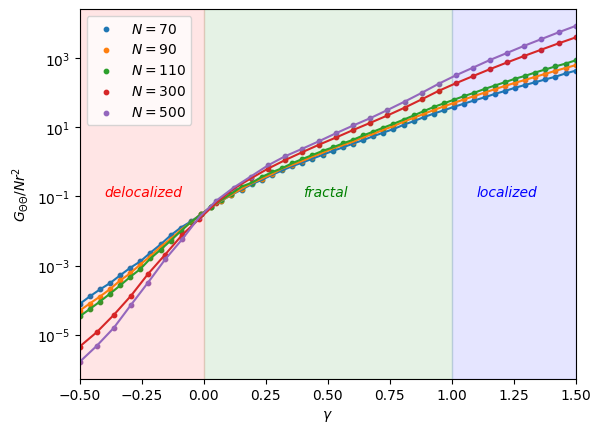} \\ }
\end{minipage}
\hfill
\begin{minipage}[h]{0.49\linewidth}
\center{\includegraphics[width=1\linewidth]{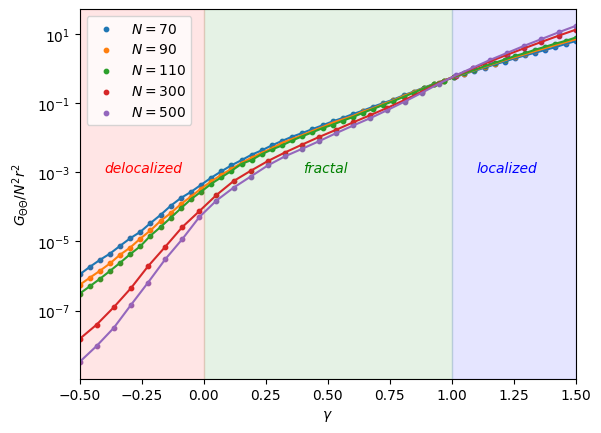} \\}
\end{minipage}
\caption{$G_{\theta\theta}$ for $r = 1/N^{\gamma}$ for different $N$}
\label{fig:G_tt}
\end{figure}

\begin{figure}[H]
\begin{minipage}[h]{0.49\linewidth}
\center{\includegraphics[width=1\linewidth]{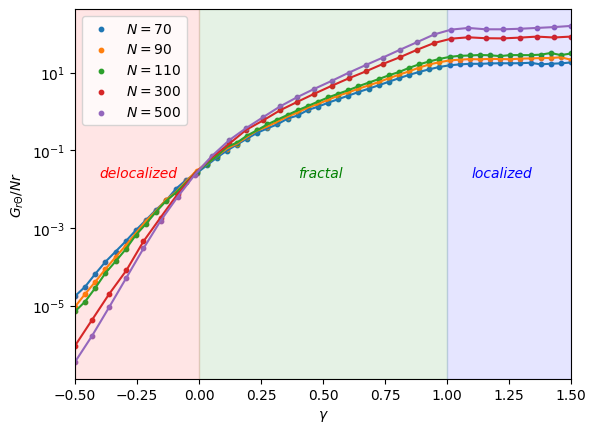} \\ }
\end{minipage}
\hfill
\begin{minipage}[h]{0.49\linewidth}
\center{\includegraphics[width=1\linewidth]{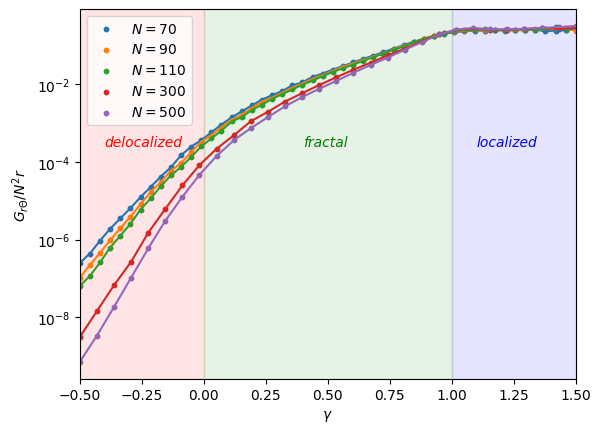} \\}
\end{minipage}
\caption{$G_{r\theta}$ for $r = 1/N^{\gamma}$ for different $N$}
\label{fig:G_rt}
\end{figure}
Our numerical study shows that there are three different regimes for $G_{\alpha\beta}$ separated by fixed point at $\gamma = 1$ with $G_{rr}/N^2$, $G_{\theta \theta}/(rN)^2$, $G_{r\theta}/(rN^2)$ independent of $N$ and at $\gamma = 0$ with $G_{rr}/N$, $G_{\theta \theta}/(r^2 N)$, $G_{r\theta}/(rN)$ independent of $N$. Similar fixed points were observed in the model discussed in \cite{sharipov2024hilbert}. Remark that in \cite{sharipov2024hilbert} at the fractal - localized transition point there is tangency of curves rather than intersection.

 In the localized phase we can use two-level approximation to obtain metric. For $G_{rr}$ and $G_{\theta\theta}$ the two-level contribution should be multiplied by factor $N(N-1)$, because every summand in the definition of $g$ (\ref{g_def}) has the same distribution. 
\begin{equation}
    G_{rr} = \frac{\pi (N-1)}{2rW}
\end{equation}
\begin{equation}
    G_{\theta\theta} = \frac{\pi r (N-1)}{W} 
\end{equation}
These arguments are in perfect agreement with the numerical results shown in Fig.\ref{fig:G_rr}. For $G_{r\theta}$, the two-level contribution vanishes, and a three-level approximation must be used instead.

To summarize, in this section we have derived asymptotic expressions for the QGT in the localized phase, which arise from the effective two-level contribution for different disorder strengths. Geometrically, this asymptotic behavior corresponds to the embedding of a two-dimensional manifold into $\mathbb{R}^3$. Depending on the ratio of the distribution width to the level spacing, $W/\delta$, the embedding exhibits either the topology of a sphere or a cone. However, for the deterministic model and the weak-disorder case, the metric does not exhibit a clear phase identification, as seen in Fig.(\ref{QGT_det}) and Fig.(\ref{QGT_weak}), in contrast to the strong disorder case shown in Fig.(\ref{fig:G_rr}), Fig.(\ref{fig:G_tt}) and Fig.\ref{fig:G_rt}.
\section{ RDM, vortex strings and BPS fractality in ${\cal N}=2$ $N_F=2N_C$ SQCD.}

\subsection{Vortex string in SQCD and BA equations}

In the previous sections, we argued that the fractality in RDM can be derived from the BA equations in the one-Bethe root sector. Since the BA equation in RDM coincides with those for twisted inhomogeneous XXX chains, we shall exploit the appearance of these spin chains in the description of the Hilbert space in the worldsheet theory on
the surface operators or semilocal vortex strings in ${\cal N}=2$ SQCD at $N_F=2N_C$. In ${\cal N}=2$ SQCD according Bethe/gauge duality  BAE yields the vacua in the vortex string
worldsheet theory \cite{nekrasov2009supersymmetric,nekrasov2010quantization,dorey2011quantization} and provides the masses of the BPS states.
We will use our findings in RDM to  argue that an analog of the fractality  and fortuity phenomena takes place in the  sector of ${\cal N}=2$ SQCD
involving the extended observable. The sector with the single vortex string or the surface operator will be considered. The vortex string 
in the NS limit of $\Omega$-deformed theory can generically
 probe the Higgs branch \cite{nekrasov2009supersymmetric,nekrasov2010quantization,dorey2011quantization}. There is  also duality mapping the theory on the vortex string on the Higgs branch to the 4d theory without the 
vortex string under proper mapping of parameters and fluxes \cite{dorey2011quantization}. Remark that the correspondence between
the wave function of the surface defect and the wave function of the inhomogeneous spin chain has been established at an arbitrary
point at the Coulomb branch in \cite{lee2020quantum}.

The 4d ${\cal N}=2$ SQCD involves $2N_C$  matter fields in the fundamental representation of $SU(N_C)$. The vortex string at $N_F>N_C$ is semilocal and enjoys orientational and size moduli, for $N_F=2N_C$ their number is the same \cite{shifman2006non}. The worldsheet theory on  $M$ vortex strings
in the NS limit of the $\Omega$ background
is the ${\cal N}=(2,2)$ $\sigma$-model with $T^*Gr(M,N_C)$ target space
\cite{nekrasov2009supersymmetric,nekrasov2010quantization,dorey2011quantization} . The flavor fields are divided into $N_F$ fundamentals and $N_F$ anti-fundamentals with twisted masses $m_i$ and $\tilde{m_i}$, respectively.
 The complex parameter on the vortex string worldsheet reads 
\begin{equation}
    \tau_{2d} = \frac{\theta_{2d}}{2\pi} + i\chi
\end{equation}
where $\chi$ is the 2d FI term. The mapping between 4d and 2d complex couplings is as follows \cite{shifman2009supersymmetric,gerchkovitz2019new}
\begin{equation}
\chi=\frac{1}{g_{YM}^2}, \qquad \theta_{2d}=\theta_{4d} +\pi
\end{equation}
where $g_{YM}$ is the gauge coupling in the 4d theory and $\theta_{4D}$ is the  4d $\theta$ term in UV.
 
 To use the fractality of the eigenmodes we have found for RDM in SQCD we shall combine two observations.
 First, the explicit relation between the normalized partition function of the half BPS surface defect in SQCD and the
 eigenfunctions of the commuting Hamiltonians of quantum twisted inhomogeneous XXX spin chains \cite{lee2020quantum}.
 Secondly, the identification of the RDM Hamiltonian as one of the spin chain Hamiltonians \cite{bork2015particle}. Combining
 these facts, we shall conjecture the fractal behavior of the particular corresponding gauge theory observables.

 To recall the relation between SQCD and spin chains, first recall the Bethe/gauge correspondence \cite{nekrasov2009supersymmetric,nekrasov2010quantization}.
 The extremization of the effective twisted superpotential in 2d theory ${\cal W}_{2d}$ yields the BA equations in worldsheet theory.
 \begin{equation}
    e^{-2i\tau_2}\prod_{l = 1}^{N_C} \frac{E_a - m_l - i\omega}{E_a - \tilde{m_l} + i\omega} = \prod_{b=1}^M \frac{E_a-E_b - 2i\omega}{{E_a- E_b + 2i\omega}}
\label{bae4}
\end{equation}
 The parameter $\omega$ in the BAE which plays the role of the Planck constant in the inhomogeneous XXX spin chain is the angular velocity in the $\Omega$ background  in the Nekrasov-Shatashvili limit \cite{nekrasov2009supersymmetric,nekrasov2010quantization}
The sector of the BAE with the $M$  Bethe roots corresponds to the M vortex strings. The 2D-4D correspondence in the
Nekrasov-Shatashvili limit of $\Omega$-deformed theory has been discussed in \cite{dorey2011quantization,chen2011new}
at the quantization locus of the Coulomb branch. The 2d and 4d superpotentials $W_{2d},W_{4d}$ coincide after the proper identification 
of parameters,
The parameter $x$ plays the role of the chemical potential or
equivalently the general shift of the twisted masses.

To obtain the identification of eigenmodes at the arbitrary point of the Coulomb moduli space, one defines the
normalized surface defect partition function \cite{lee2020quantum}
\begin{equation}
    \Psi(\hbar)= \lim_{\omega_2\rightarrow 0}\frac{Z_{defect}(\omega_1,\omega_2)}{Z_{bulk}(\omega_1,\omega_2)}
\end{equation}
$\hbar =\omega_1)$, which in the NS limit behave as 
\begin{equation}
    Z_{defect}=e^{\frac{\cal{W}}{\omega_2}}(\Psi(\hbar) +\dots)
\end{equation}
\begin{equation}
    Z_{bulk}= e^{\frac{\cal{W}}{\omega_2}}
\end{equation}
It was shown in \cite{lee2020quantum} that the quantum Hamiltonians of the spin chain are explicitly related 
to the commuting subalgebra of observables   in gauge theory and the normalized partition function
of the surface defect at any point of the Coulomb moduli is proportional to the eigenfunction of
the spin chain transfer matrix. We use this correspondence at the quantization locus where $W_{2d}=W_{4d}$.

Consider now the second ingredient, the relation of the RDM model and the inhomogeneous spin chain supporting its integrability, 
first observed in \cite{Dunning2004integrability} at the BAE.level. The identification of RDM in the spin chain
context has been done in \cite{bork2015particle}. 
Recall that the quantum XXX chain is described by the monodromy matrix (see Appendix)
 $\cal{T}(\lambda,\varepsilon_i,\theta)$ 
 which upon taking trace over the auxiliary space yields the 
 transfer matrix
 \begin{equation}
     T(\lambda,\varepsilon_i,\theta)= Tr_{aux} \cal{T}(\lambda)
 \end{equation}
 The transfer matrix  is the generating function for the commuting nonlocal Hamiltonians 
 of the inhomogeneous XXX spin chain
 $ T(\lambda,\varepsilon_i,\theta)=\sum_{k=1}\lambda^{-k} t_k$.
 Remarkably \cite{bork2015particle} the RDM Hamiltonian is identified with the first nontrivial Hamiltonian of the spin chain
 \begin{equation}
     H_{RDM}\propto t_2
 \end{equation}
 which explains the identity of the BAE.

To obtain the mapping of parameters in RDM and SQCD compare the BA equations for the vortex string (\ref{bae4}) and for the RDM model.

 \begin{equation}
    e^{-2i\theta}\prod_{l = 1}^{N} \frac{E_a- \varepsilon_l - ir\sin\theta}{E_a - \varepsilon_l + ir\sin \theta} = \prod_{b=1}^M \frac{E_a-E_b - 2ir\sin \theta}{{E_a- E_b + 2i r\sin \theta}}
\end{equation} 
First, remark that the sector with M vortex strings corresponds to the sector with M Cooper pairs in RDM. The parameters $\varepsilon_i$ correspond to the 2d twisted masses that are assumed to be real in RDM. The twist in the RDM model $\tau_{RDM}$ is real, hence 
\begin{equation}
\chi=\frac{1}{g_{YM}^2}=0
\end{equation}
and we are at the strong coupling point in 4D SQCD. The $\theta$ parameters are related as follows,
\begin{equation}
    \theta_{RDM}= \theta_{4d}-\pi
\end{equation}
hence, the Richardson limit corresponds to $\theta_{4d}\rightarrow \pi$. The RDM BA equations imply the relation $m_i=\tilde {m}_i$
for the fundamental and antifundamental masses. The relation between the masses of fundamentals and antifundamentals implies that 
all spins in the chain are the same $s_1=1/2$. 
 
The unusual point is that the effective chemical potential $x$, the twist $\theta$ and the parameter of $\Omega$-deformation $\omega$ are related.
\begin{equation}
    x=r\cos \theta ,\qquad \omega= r \sin\theta
\end{equation}
Generically,  Planck constant in the XXX spin chain, the parameter of $\Omega$-deformation, and the chemical potential 
are independent parameters. 
However, in the RDM case $\Omega$-deformation appears to be an emerging phenomenon of the combined effect of the chemical potential and the $\theta$-term. 
In summarizing, we get the relation between the eigenmode of the RDM Hamiltonian in one Bethe root sector and the defect wave function
, which allows us to conjecture the fractality properties of the latter.

 It is worth commenting on the brane picture behind our strong coupling limit in the IIA setup \cite{Witten_1997}. The brane configuration 
 for $N_F=2N_C$ SQCD involves two parallel NS5 branes at distance $\frac{1}{g^2}=\delta x_6$ in one direction, N D4 branes suspended between them and N semiinfinite D4 branes  both on the left and right sides of the left and right  NS5 branes. Naively, there is strong coupling U(2) symmetry 
 at the $\frac{1}{g^2}= 0$ point we are working with; however, it is broken by non-vanishing $\theta$-term. 
 Since $m_i=\tilde{m}_i$ the semi-infinite D4 branes on the left and right join together.
 The positions of 
 D4 branes  correspond to the quantization locus \cite{dorey2011quantization, Jeong_2021} on the Coulomb branch,  therefore the Higgs branch opens 
 and we can move one NS5 brane apart. The  D2 probing the Higgs phase yields the  vortex string or surface defect if its tension tends to infinity..
 The BAE we are looking at corresponds to the extremization of the effective twisted superpotential in the worldvolume theory at the 
 single D2 brane. The 1/4 BPS monopole-kink states on the surface defect are represented by the additional 
 properly oriented D2 branes \cite{hanany2003vortices}.

\subsection{ BPS chaos and BPS fractality}

Having formulated the relation between the specific subsector
of ${\cal N}=2$ SQCD and a single Bethe root solution of the RDM model, we can now conjecture the notion of BPS fractality generalizing the discussion in \cite{chen2025bps} where
the notion of BPS chaos was introduced.
It was motivated by the search for the mechanism responsible for formation of a black hole horizon.  The Bekenstein-Hawking entropy is believed to be saturated by the ensemble of degenerate BPS states, which is generally assumed to have a small $\frac{1}{16}$ amount of SUSY. Therefore, it is natural to question the difference between the properties of microstates that form the BH horizon and those that form only horizonless geometries.

There are two ways to approach this problem.
First, it was suggested that the chaotic properties of the microstates play a key role. Usually, chaotic properties are quantified by the spectral statistics; however, in the horizon case, the situation is different, since it involves the multiple degenerate states whose multiplicities produce the required entropy. Therefore, different diagnostics of the chaos in the degenerate spectrum is required.  It was suggested to consider some probe operator with nontrivial matrix elements between degenerate states and investigate its spectrum \cite{lin2023holography}. The statistical properties of this operator allow us to say if the sector of the Hilbert space is chaotic or not. The statistical properties were investigated via the simplest diagnostics, the level spacing distribution. The Wigner surmise corresponds to "BPS chaos", while deviation in the direction of the Poisson distribution leads to "BPS localization".

Testing of this line of ideas in the context of manifolds of BPS states with different amounts of SUSY shows that it works reasonably well \cite{chen2025bps}.
For states with $\frac{1}{2}, \frac{1}{4}, \frac {1}{8}$ SUSY the chaos is weak enough, while for BPS states with $\frac{1}{16}$ SUSY which are assumed to be responsible for the BH entropy the chaoticity of the spectrum is found. The choice of operator was a subtle issue, and only a few examples have been suggested in \cite{lin2023holography,chen2025bps}.
It was argued that the band structure of the matrix representing the operator on the basis of the vacuum states is essential in chaotic properties.

The second direction has the holographic origin and deals with the question of what are the properties of
large N gauge dual theory theory which allow or not allow the horizon to form in the dual geometry. It was argued that gravitons correspond to the operators in the dual gauge theory which go smoothly in the $N\rightarrow \infty $ limit, while to get the horizonful dual geometry one has operators which undergo some transitions at finite N. This behavior has been
formulated in terms of the R-charge concentration and fortuity \cite{chang2023words, chang2024decoding,chang2024holographic,chang2024fortuity}. Roughly speaking, it claims that operators are BPS at fixed N only at some values of the some global charge, usually R charge. Another term used in the same context is the "BPS invasion" \cite{chen2025bps}, which is the counterpart of the particular wall-crossing phenomenon on the parameter space.

The BPS states are represented by proper cohomologies; hence these arguments have a precise mathematical counterpart found in \cite{choi2024finite}. It was found that two types of cohomologies have to be distinguished ; ones which are smoothly extended in the $N\rightarrow \infty$ limit and the second type that involves some additional sensitivity to finite N. Physically, the properties at large N are important in the horizon context, which implies some restriction of N to form the horizon. For example, in some cases it can be reformulated as the string exclusion principle \cite{maldacena1999ads3}.

Let us conjecture that the Hamiltonian of the RDM model plays the role of the
probe of the specific BPS sub-sector in ${\cal N}=2$ SQCD. 
\begin{equation}
    H_{ij}=\langle i|\hat{H}_{RDM}|j \rangle
\end{equation}
We can identify both the chaoticity of the specific states and the nontrivial dependence on N.
The effective parameter in our case is $(\gamma \log N)$, so it is useful to follow the dependence on $\gamma$.

Our analysis in the previous sections can now be interpreted as an investigation of the  fractality and stability of charge Q states in the domains in the $(\gamma \log N, \theta)$ parameter space. The domains
are separated by the CMS that we have identified numerically. We emphasize that, contrary to the analysis of CMS in \cite{ievlev2020string}
where $\Im \tau_{2d}\neq 0, \Re \tau_{2d}\neq 0, r=0$
we consider CMS in the  $(\Re\tau_{2d} \neq 0, r\neq 0)$ parameter plane  assuming 
$\Im \tau_{2d}=0,$.  The fractality found in the single magnon sector
now corresponds to the fractality of the BPS states with particular 
values of the global charge.

Remark that $H_{RDM}$, which serves as a probe of fractality and chaoticity
in the sector of the Hilbert space of ${\cal N}=2$ SQCD
has similarity to the fermionic operators suggested in \cite{chen2025bps} for the same purpose.
Here, instead of the single fermion we consider the Cooper pair. Moreover,
contrary to \cite{chen2025bps} we consider long-range hopping in the sector
of the Hilbert space under consideration
and, equivalently, elaborate deterministic or disordered versions of the RDM model. 
Additionally, the cyclic RG where $\log N$ plays the role of time shows that the
limit $N\rightarrow \infty$ in this subsector of BPS states is not smooth.

Hence we conjecture that the chaotic and fractal properties of RDM model capture the
fractal properties of the single vortex string/surface operator BPS sector of ${\cal N}=2$ SQCD. We have considered the diagnostics of the eigenfunctions instead of the diagnostics of the spectrum. It is more useful and sometimes more informative. Our tools are fractal dimensions, quantum metrics, and the value of charge $Q$. All three characteristic uncover the presence of three distinct phases and two of them allow to identify them as localized, multifractal, and delocalized phases.
Moreover $\langle Q \rangle$ allows us to recognize the detailed structure of the multifractal phase. 
We identified a few elements that were very essential in the analysis in \cite{chen2025bps}. First, we have seen that the departure into the multifractal phase that changes the value of $\gamma$ is signaled by the change of the band structure in the matrices. Secondly, we have clearly identified that in the regime of BPS fractality there are multiple lines of instability of BPS states, which is the direct analog of the "invasion of non-BPS states" is the spectrum in \cite{chen2025bps}. 

Let us emphasize that our claim about BPS fractality is based on the exact solution of the BAE in one root sector and the 
identification of an RDM Hamiltonian with one of the non-local Hamiltonians of the inhomogeneous XXX spin chain governing
the surface defect/vortex string wave function. The very fractality in the BPS sector involve a few key ingredients
\begin{itemize}
    \item Effective modification of the plane wave solution anzatz for BAE which is in agreement with a similar modification 
underlying the fractality as discussed in  \cite{monthus2017multifractality} 
    \begin{equation}
        B(E)\ket{\rm{vac}}=\sum_i\frac{b^+_i}{E-\varepsilon_i}\ket{\rm{vac}} \rightarrow \sum_i\frac{e^{-i(\varphi_{j} + \varphi_1+ 2\sum_{k = 2}^{j-1} \varphi_k)}b^+_i }{\sqrt{(E-\varepsilon_i)^2 +y^2}}\ket{\rm{vac}} 
    \end{equation}

This Breit-Wigner modification implies a kind of instability which in the context of the vortex string 
means the appearance of the electric fluxes $Q_i$ whose number depends on the scaling parameter $\gamma$. 
Physically such instability due to flux formation could have two origins-Schwinger pair production
since $\theta$ is an effective electric field or  the Witten effect of the emergent electric charge
of kink-monopoles in the 2d theory with $\theta$-term
\item The nontrivial N scaling of the Planck constant in the spin chain context which has the meaning
of the angular velocity in $R^4$ or equivalently the equivariant parameter in the instanton counting problem
\end{itemize}
We have made only the first step in the identification of the BPS fractality phenomenon, and certainly 
additional analysis is required to analyze other BPS sectors of the model.

Let us comment on the cohomological counterpart of our findings.
The BAE for RDM provides the description of the vacua and excitations
in the vortex string sigma model and represents the quantum cohomology ring of its target manifold $T^*Gr(K,N)$ for K vortex strings
\cite{nekrasov2009supersymmetric,nekrasov2010quantization} and the parameter of $\Omega$-deformation 
plays the role of an equivariant parameter. If we consider the single vortex string, it corresponds to the excitation with the single Bethe root
and the target manifold is $T^*CP(N-1)$. The classification of the BPS states representing quantum cohomologies is quite rich even for the compact target manifold $CP(N-1)$ \cite{dorey1998bps,dorey1999bps} without equivariant parameter. 
There are $N$ vacua and the 1/2 BPS kinks interpolate between them. Their masses are determined by the central charges in the SUSY algebra and
are fixed by the differences of the twisted superpotentials in the two vacua.
The kinks with the different global charges decay at the curves of marginal stability, which form the different patterns
on the manifold of twisted masses.

Our case is a specific limit of the cohomological problem. We consider the BAE which governs the quantum equivariant
cohomology ring for $T^*CP(N-1)$. The quantum parameter is the pure phase $q=e^{i\theta}$, the twisted masses are real and
the equivariant parameter for $U(1)$ rotation in $\Omega$ deformation is related to the quantum parameter $\omega\propto \sin\log q$.
We have found the wall-crossing family for elements in the ring that decay at different values of $\log q$,
which is the counterpart of the BPS-invasion phenomenon. It
would be very interesting to formulate the BPS fractality purely in the cohomological language.

\subsection{Comments on CFT representation of eigenfunctions, Matsuo-Cherednik duality and brane picture}

 Let us make some brief comments concerning the geometrical aspects of the problem postponing the detailed discussion to a separate study. We combine some arguments based on the CFT representation of the
eigenfunctions of Gaudin and inhomogeneous XXX spin chains, the classical-quantum version of the Matsuo-Cherednik duality and brane picture.
The mapping to the Calogero and Ruijsenaars models is fruitful because they represent a universal class of systems
with anyonic statistics and provide additional intuition.

Underlying geometry is the two-dimensional plane in the Higgs branch which can be considered
as the section in the 4d manifold where the auxiliary gauge theory with flavor group is defined. 
The geometrical framework for BAE for the Gaudin model underlying Richardson has been developed in \cite{gaiotto2012knot} in the context of knot homologies. It was argued that one has to consider the 4d gauge theory
on the manifold $M_3\times R_{+}$ where the knot is implemented in the boundary manifold $M_3$. It was argued that the Gaudin model describes the interaction of two types of magnetic objects. The "singular" fixed monopoles are located at points $\varepsilon_i$ on the 2d slice of the 4d manifold, whereas
the Bethe roots correspond to the "movable" magnetic defects. The knot can be considered as braiding of the movable magnetic object around the fixed
"singular" monopoles.
The operator generating an irregular singularity at infinity corresponds to vev of the scalar responsible for the symmetry breaking of the $SU(N)$ gauge.
This gauge theory for the Gaudin model can be interpreted as the Hitchin model on the sphere \cite{nekrasov1996holomorphic} however to obtain the inhomogeneous twisted XXX spin chain the
topological-holomorphic 4d Chern-Simons theory is more suitable \cite{costello2020unification}.

This geometry fits our consideration as follows. The singular fixed monopoles correspond to the positions of the flavor branes
at $m_i$ while the "movable" magnetic object corresponds to the D-brane representing the magnetic string populated with confined monopoles.
Since we are at the strong coupling point $\frac{1}{g_{YM}^2}=0$, the NS5 branes coincide, yielding the non-perturbative $SU(2)$.
To introduce the $\theta$ term, one
takes into account the $U(1)$ gauge field in the RR sector with curvature 2-form $f=da$ with non-vanishing flux over the disc \cite{witten1998theta}
\begin{equation}
    \int_D f=\theta +2\pi k
\end{equation}

We expect on the 2d plane, where the radial coordinate can be thought of as the Liouville radial
coordinate in the holographic setting, the following ingredients behind the XXX and RDM BAE. 
First, the insertions of the local operator at the points $m_i=\varepsilon_i$, the nonlocal operator representing
the vortex string, the flux of the RR field f providing the dependence $\theta$, and the background radial U(1) field
corresponding to the chemical potential. The CFT representation for the Gaudin-Richardson case
was well developed \cite{sierra2000conformal, gaiotto2012knot,biskowski20252dcftefficientbethe}
and we recall its ingredients below, while the CFT representation of the RDM-XXX spin chain is more complicated,
and we refer the reader to \cite{lee2020quantum,nekrasov2022surface,Jeong_2021}.

To get the CFT representation for the Gaudin-Richardson case,
consider the perturbed chiral conformal block in the theory with central charge. 
\begin{equation}
    c=1+Q^2,\qquad Q=b+b^{-1}
\end{equation}
 chiral vertex operators $V_{\alpha}= :e^{2\alpha \phi(z)}:$ of conformal weight
\begin{equation}
    \Delta_{\alpha}=\alpha (Q-\alpha)
\end{equation}

The conformal block for the Richardson model involves the operators corresponding to Bethe roots $E_i$, inhomogeneities $\varepsilon_i$, and twist G \cite{sierra2000conformal,biskowski20252dcftefficientbethe}. 
\begin{equation}
     <\prod_{i=1}^N \Psi_{(2,1)}(\varepsilon_i)\prod_{i=1}^M V_{\frac{1}{b}}(E_i)V_G >
\end{equation}
The $\Psi_{(2,1)}$ is the degenerate field that yields the regular singularities at the points $\varepsilon_i$ and $V_G$ amounts to the irregular singularity at infinity corresponding to twist. The operators $V_{\frac{1}{b}}(E_i)$ correspond to the Bethe roots or equivalently to a single Cooper pair and are interpreted as screening operators in the CFT framework \cite{sierra2000conformal}. The operators $V_G$ at infinity break down the conformal invariance
and correspond to the Gaiotto vector, which is the eigenvector of the $L_1$ Virasoro operator

The relevant solution to the KZ equation is as follows.
\begin{equation}
    \Psi_{KZ}(\varepsilon_i)= <\oint dE_1\dots\oint dE_M \prod_{i=1}^N \Psi_{(2,1)}(\varepsilon_i) \prod_{i=1}^M V_i(E_i)V_G>=
    <\oint dE_1\dots\oint dE_M  e^{-\frac{1}{b}W(\varepsilon_i, E_i)}> 
\end{equation}
where the Yang-Yang function for the Gaudin model under consideration is read as 
\begin{equation}
    {\cal W}(\vec{\varepsilon}, \vec {E})= \frac{1}{2} \sum_{i<j} \log(\varepsilon_i-\varepsilon_j) +2 \sum_{\alpha <\beta} \log(E_{\alpha}- E_{\beta})
    - \sum_{i}\sum_{\beta} \log(E_{\beta}-\varepsilon_i) +\frac{1}{g} (-\sum_i\varepsilon_i +2\sum_{\beta} E_{\beta})
\end{equation}
see, for instance,  recent discussion in \cite{biskowski20252dcftefficientbethe}. 
The semiclassical limit $b\rightarrow 0$ results in the BA equations. 
\begin{equation}
    r_i=\frac{\partial W}{\partial \varepsilon_i}
\end{equation}
where $r_i$ are the eigenvalues of the Gaudin Hamiltonians $R_i$.

According to Matsuo-Cherednik duality, the variables $\varepsilon_i$ in the XXX chain
in the dual many-body integrable rational RS model become the coordinates. 
At the quantum-quantum level duality relates the solutions to the KZ or qKZ equations and the totally symmetric or antisymmetric wave functions of the
Calogero-Ruijsenaars-Schneyder(RS)-Toda family of integrable models \cite{matsuo1992integrable, cherednik1994integration,givental1995quantum}. 
However, we are interested in the so-called classical-quantum limit when this duality reduces to the relation between the BA equations in the inhomogeneous spin chains and the intersection of the Lagrangian submanifolds in the classical RS models. The relevant classical-quantum pair in our case is the inhomogeneous quantum twisted XXX spin chain with which we worked 
and the rational RS model \cite{gorsky2014spectrum}. Different aspects of the Matsuo-Cherednik duality and the complete list of references can be found in \cite{gorsky2022dualities}.

 The explicit relation between XXX and the rational RS model is as follows. We define the transfer matrix of the inhomogeneous GL(n) XXX spin chain at N sites depending on the formal spectral parameter which serves the generation function for the non-local Hamiltonians $H^{XXX}_j$.
 \begin{equation}
     T^{XXX}(z)=TrV + \sum_{j=1}^N \frac{H^{XXX}_j}{z-x_i}
 \end{equation}
 where V is the GL(n) twist matrix $V=diag(V_1,\dots,V_n)$. The eigenvalues of $H^{XXX}_j$ depend
 on the inhomogeneities $x_i$ and on the solution of the system of the BA equation, which are nested in the generic case 
 $$(\{\mu_i^1\}_{N_1}, \dots, \{\mu_i^{n-1}\}_{N_{n-1}})$$
where $N_a$ denotes the number of Bethe roots at the a-th level of the nested BA. There is no nesting for the GL(2) case we are working with.

 In the dual rational RS model, we define the Lax matrix
 \begin{equation}
     L^{RS}_{ij}= \frac{\hbar \dot{x_j}}{x_i-x_j +\hbar}, \dots i,j=1\dots N
 \end{equation}
 which yields the RS Hamiltonian, which is
 \begin{equation}
     H^{RS}=TrL^{RS}= \sum_{j=1}^Ne^{\eta p_j}\prod_{i\neq j}^N
     \frac{x_j-x_i+\eta \nu}{x_i-x_j}
 \end{equation}

 According to duality 
 \begin{equation}
    \dot{x_j}=\frac{1}{\hbar}H_j^{XXX}(x_j,\mu_j) 
 \end{equation}
and the eigenvalues of the RS Lax operator have non-trivial multiplicities $\mathrm{mult}V_1=N-N_1, \mathrm{mult} V_2=N_1-N_2,\dots, \mathrm{mult} V_n=N_{n-1}$ \cite{gorsky2014spectrum}.
 
 The Planck constant in the spin chain $\hbar_{XXX}= r\sin\theta$ is assigned to the parameters of the rational RS model.
\begin{equation}
    \varepsilon_i\leftrightarrow x_i, \quad \hbar_{XXX} \leftrightarrow \nu \eta , 
\end{equation}
where $\eta$ is the relativistic parameter and $\nu$- is the coupling constant.
 Hence for GL(2) case we have for $\eta=1$ the RS particles interacting 
 with coupling $\nu=r\sin \theta$
 When $\nu \rightarrow 0$ the classical-quantum duality relates the 
 Gaudin model and the rational Calogero model in the same manner.

 Hence in the Gaudin-Richardson limit in our problem we obtain the rational Calogero model of particles at $\varepsilon_i$ with interaction $\frac{1}{(\varepsilon_i - \varepsilon_j)^2}$.
Since the rank of the group $GL(2$) does not coincide with the number of sites of the Lax operator, the eigenvalues are degenerate and form two groups.
The Calogero particles are located along the radial
Liouville coordinate.

 In the RDM-XXX case the situation is more tricky. We can consider 
 the rational RS model with $\eta \nu = r\sin \theta$ or 
the trigonometric Calogero model related to the rational RS model via bispectral duality \cite{fock2000duality}. In the bispectral dual Calogero 
 model, the clustering of the eigenvalues of the RS Lax matrix gets mapped into the clustering of Calogero particles on the circle into two groups.
 Remark that the relevant geometry is three-dimensional and involves
 $(r,\phi,\tau_E)$ coordinates where $\tau_E$- is the Euclidean time direction.
 There are fluxes of the $U(1)$ gauge field $F_{\tau r}=\rho$ due to the chemical potential and density
 $A_\tau (r\rightarrow 0)= \mu + F_{r\tau} r$ and flux of the RR 2-form
 $f_{r\phi}$ due to the $\theta$-term. It would be interesting to compare this setup with the appearance of the Calogero and Ruijsenaars models as boundary excitations in FQHE.

\section{Discussion  }
\subsection{Comments on the Luttinger-Ward relation}
We compare the results for the global charge Q obtained by BA with the somewhat similar relations for the global charge in a
complex SYK model.
The conventional SYK model at the low-energy limit is dominated by the
Goldstone mode from the breaking of the diffeomorphism invariance of Euclidean time. The complex version of the SYK model \cite{gu2020notes,davison2017thermoelectric} involves the second degree of freedom $\phi(\tau)$ - phase of the complex fermion,

First, we just consider the complex SYK model with spectral asymmetry.
Remarkably, the complex SYK with spectral asymmetry has a holographic interpretation as the near-horizon region of the higher-dimensional charged black hole which has $AdS_2\times\ R^{d-2}$ near-horizon geometry. The spectral asymmetry parameter $\theta$ has the meaning of the near-horizon electric field, while the parameter $\langle Q\rangle$ is interpreted as the asymptotic charge density \cite{sachdev2015bekenstein}. 
The nonzero entropy of the dual extremal charged
BH at zero temperature obeys the following relation.
\begin{equation}
    \frac{dS({\cal Q})}{d{\cal Q}}=2\pi {\cal E}
\end{equation}
where $S({\cal Q})$ is the mean entropy of one degree of freedom and 
\begin{equation}
    Q_{SYK}=\sum_i b_i^{\dagger}b_i -\frac{N}{2} = N \cal{Q}
\end{equation}
is the fermion number or equivalently the R-charge.
In the holographic dual the zero-temperature entropy is interpreted as
either the area in the higher-dimensional sphere in the near horizon limit or as an effect of the massive bulk fermions in $AdS_2$ in the
background electric field \cite{gu2020notes}. 

The particle hole symmetry breaking parameter ${\cal E}$ 
in complex SYK
can be introduced in both UV and IR regimes. In UV it is defined via the fermion Green function as 
\begin{equation}
    G(\tau_1,\tau_2)=-\langle T\Psi^\dagger(\tau_1){\Psi}(\tau_2)\rangle,\quad 
    G(0^+)=-\frac{1}{2} + {\cal Q}, \quad  G(0^-)=\frac{1}{2} + {\cal Q}
\end{equation}
On the other hand, in IR the asymmetry is introduced via
small frequency behavior of the fermion Green function \cite{gu2020notes}
at $\beta\rightarrow \infty $
\begin{equation}
    G(\omega)_{\omega \rightarrow 0}(\pm i\omega)\propto \pm e^{\mp i\theta}
    \omega^{2\Delta -1}
\end{equation}
 - where $\Delta=1/q$ is the classical scaling dimension of the fermion in the qSYK model upon rescaling of time. Another suitable parameterization of the particle-hole breaking parameter
is provided below. 
\begin{equation}
    e^{2\pi {\cal E}}=\frac{\sin(\pi \Delta +\theta)}{\sin(\pi \Delta -\theta)}.
\end{equation}
In this parametrization condition of real $\cal E$ implies restriction on $\theta$: $\theta \in (-\pi\Delta,\pi\Delta)$ - region of unitarity of the model.

Let us compare the BAE in RDM with similar relations for the
global charges in complex SYK \cite{gu2020notes} and ${\cal N}=2$ \cite{turiaci2023mathcal,heydeman2023phases}. In that case,
the corresponding formulae were identified as the Luttinger-Ward relations.
The SYK is the popular model of the quantum dot with the long-range interaction \cite{sachdev1993gapless,kitaev2015simple}
and is considered a candidate for the description of microstates of the BH horizon.
The Luttinger-Ward relation between $\langle Q \rangle$ and $\theta$ found in \cite{gu2020notes} is as follows:
\begin{equation}
    {\cal Q}=\frac{\langle Q_{SYK}\rangle}{N}=-\frac{\theta}{\pi}+ (\frac{1}{2} -\Delta) \frac{\sin 2\theta}{\sin 2\Delta}
    \label{qterm}
\end{equation}
where $\Delta=1/q$ is the classical scaling dimension of the fermion in the qSYK model upon the rescaling of time.  In \cite{gu2020notes} it was argued
that the second term in (\ref{qterm}) follows from a kind of anomaly and reflects the correlation of the UV and IR scales. Alternatively, it can be obtained from the ratio of 2D fermion determinants with the particular boundary conditions \cite{gu2020notes} and to some extent it can be interpreted as the renormalization of $\theta$ by the fermionic loops.

There is a clear similarity between the Luttinger-Ward relation (\ref{qterm}) and our result for Q  in the single pair sector
$$
    \langle Q\rangle  = \frac{\theta}{\pi} + \frac{1}{\pi}\sum_{l=1}^N\langle \arctan{\frac{y}{(E-x - \varepsilon_l)}}\rangle_{\varepsilon}
$$
The parameter $\theta$ in \cite{gu2020notes} was introduced as the twist of the fermion Green function in $\omega\rightarrow 0$, while in our case it is introduced similarly for the Green function of the Cooper pair. Since our Hamiltonian is quadratic, the second term can also be interpreted as the effect of determinant, which is now bosonic. 
\begin{figure}[H]
\begin{minipage}[h]{0.49\linewidth}
\center{\includegraphics[width=1\linewidth]{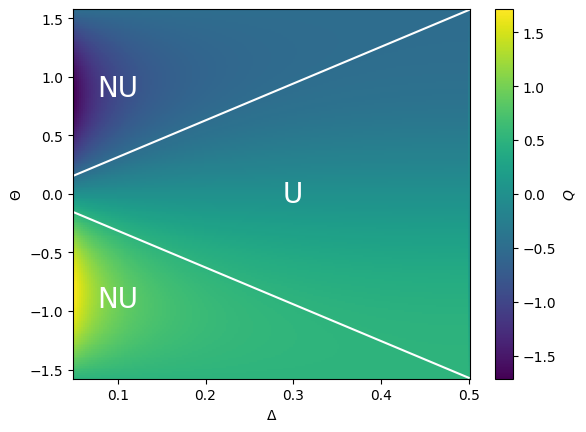} \\ }
\end{minipage}
\hfill
\begin{minipage}[h]{0.47\linewidth}
\center{\includegraphics[width=1\linewidth]{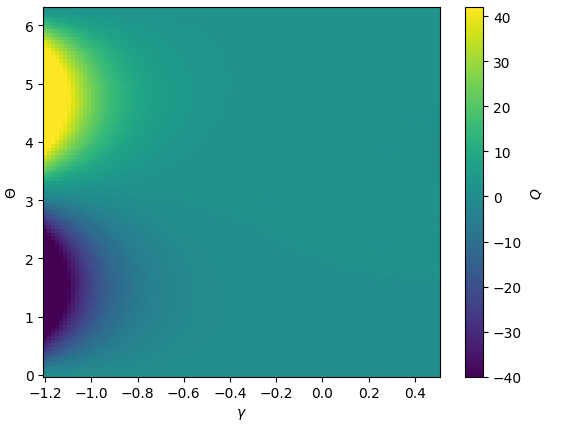} \\}
\end{minipage}
\caption{Comparison ${\cal Q}$ dependence of $(\Delta, \theta)$ for and $Q$ for clean RDM. U stands for unitary region, NU - non-unitary.}
\label{ris:image1}
\end{figure}

Hence, it is natural to question whether there is an analog of the concentration of global charge in the complex SYK with spectral asymmetry
similar to the RDM case. 
The parameter plane $(\gamma,\Delta)$ for SYK has the same high and low charge regions as in the RDM, but in the case of complex SYK transition occurring in a non-unitary region (NU), see Fig.(\ref{ris:image1}). In the RDM case, charge exhibits a multifractal region, while in SYK there are no "chaotic" regions for $\theta \in (-\pi \Delta, \pi \Delta)$.
It would be interesting to add the complex SYK Hamiltonian with spectral asymmetry to the RDM model and investigate the transition from the strange metal to the superconducting state. Or, in contrast, add the attractive Hubbard term to the complex SYK with the spectral asymmetry generalizing the analysis in \cite{wang2020sachdev}. From our analysis we know that RDM involves rotation and the $\theta$-term from the 4-dimensional perspective, hence it cannot be
obtained by simple interpolation from SYK with broken particle-hole symmetry.

Additionally, we expect that the pseudogap phase takes place between perturbed SYK and RDM regimes
with non-trivial out-of-equilibrium phenomena similar to ones discussed in the SYK + U model in \cite{alexandrov2023out}. This line of research promises an interesting
outcome for the near-horizon non-perturbative phenomena for the charged black holes using the complex SYK - charged BH duality. 
As we have argued, the fractality implies a kind of fragmentation of the matrices into the blocks, which in the dual gravity charged BH model
would correspond to the fragmentation of the BH. This fragmentation of the BH presumably corresponds to the partial deconfinement transition in dual large N-gauge theories \cite{hanada2019anatomy,hanada2019partial}.

\subsection{ Zero temperature entropy and dualities in the integrable systems}

Can we suggest some counterpart of the zero-temperature entropy
in the context of RDM which is discussed in the complex SYK model? To this aim,
exploit the Matsuo-Cherednik duality between the inhomogeneous spin chains
and the trigonometric Calogero model we have mentioned above.
 However, it was noted long ago that the rational Calogero model 
supplemented with an oscillator potential.
\begin{equation}
    H_{cal}=p^2 +\frac{\nu^2}{x^2}+ \omega^2 x^2
\end{equation}
describes the particle near the $AdS_2$  horizon of the extremal charged black hole \cite{gibbons1999black} where
the coordinate r corresponds to the radial coordinate in $AdS_2$. 

Clustering of particles produces degeneration of the spectrum, which qualitatively provides a microscopic description of the charged black hole entropy \cite{gibbons1999black}. 
Namely, one starts with the N -body Calogero system and considers the case when a cluster of the (N-1) particles is formed yielding the proper N scaling of the coupling constant due to a kind of falling to the center pattern. The problem of falling to the center in the Calogero model hosts the tower of Efimov states \cite{braaten2004renormalization, beane2001singular,bawin2003singular}. It goes as follows:
we introduce the cutoff in the radial coordinate and impose the proper boundary condition at the wave function. It turns out that the Calogero coupling constant gets renormalized and the tower of shallow states emerges \cite{hammer2011efimov}, this behavior can be interpreted as the anomaly in the SL(2) algebra involving the dilatation operator \cite{ananos2003anomalous, moroz2010nonrelativistic}.
The description of BH entropy through conformal quantum mechanics was made more precise in the supersymmetric case \cite{gaiotto2005superconformal, dorey2023black}
where the relevant index evaluated in superconformal quantum mechanics does the job. Therefore, we can focus on the problem
of evaluating the entropy via the Calogero model.

According to the Matsuo-Cherednik duality, the parameter $\varepsilon_i$ becomes the coordinate of RS particle $x_i$ and the Planck constant 
in the XXX spin chain becomes the RS coupling constant $\nu= r \sin \theta$
At the next step, recall  that the rational Calogero model with oscillator potential is dual to the trigonometric Calogero model \cite{nekrasov1997duality}. 
And completing the chain of dualities familiar in the many-body integrable problems recalls the bispectral Ruijsenaars duality \cite{ruijsenaars1988action} between the trigonometric Calogero model and the rational RS model, which can be interpreted in the gauge theory framework \cite{fock2000duality}. Hence, the clustering of particles in the trigonometric Calogero model corresponds to the clustering of the spectrum in the RS model.

From our study, we have seen that $Q$ effectively measures the number of "interacting" levels
in the particular domain of $(\delta, W)$, which is the counterpart of clustering in our problem.
One more remark seems to be relevant. The spectrum of Calogero model at rational coupling is related to the torus knots and torus links. The clustering of Calogero particles that is important for our problem 
corresponds to the torus link invariants \cite{etingof2015representations} and can be described in terms of representation of the DAHA spherical algebra.
In our case the coupling constant in the Calogero model dual to XXX chain is proportional to $r sin\theta$ hence we expect more clear
clustering of degrees of freedom if it would be rational, The knots can be recognized at the spin chain side as well without going to the dual Calogero model. This has been discussed in \cite{gaiotto2012knot}
for the Gaudin model, which corresponds to the Richardson model in our discussion.


\subsection{On the baryon state}

The RDM  and Richardson models describe superconductivity 
in finite-dimensional systems. Is there any counterpart in
the superconformal SQCD? To this aim, we have to recognize the 
analog of the Cooper pair and its condensation. 
Let us  take advantage of the detailed analysis
performed in \cite{ievlev2020string} for the $SU(2)$ $N_F=4$ theory, which has been elaborated near the self-dual strong coupling point.
Although the $N_C=2$ example cannot be used for the derivation of fractality, it provides some clues about the interpretation of the Cooper pair analog in ${\cal N}=2$ SQCD.

 The deformation of the self-dual point by the complex modulus was identified as the baryonic state with $Q_B=2$ built from a closed semilocal vortex string populated by four confined monopoles.  This baryonic state is massless at the $\tau_2=0$ level but decays at the marginal stability curves at the $\tau_2$ complex plane into two quarks in the bifundamental representation. The coordinate on the non-perturbative Higgs branch was identified with the baryon condensate. We emphasize that in \cite{ievlev2020string} the $\Omega$ deformation and chemical potential are absent; however, the picture supporting the baryonic excitations with $Q_B=2$ at the self-dual point is quite suggestive. The observation concerning the existence of baryonic excitation in SQCD near the self-dual strong-coupling point has also been supported by the stringy picture \cite{koroteev2016non}.
It would be interesting to identify the baryonic state in the context of superconducting interpretation of the RDM model

\subsection{Analogy with transitions in  thermal and  dense QCD}

Following the logic suggested in \cite{chen2025bps} we assumed that the RDM Hamiltonian is the operator that measures the chaoticity and fractality of the specific BPS sector in SQCD. We argue that there is an important analog of the similar phenomenon in conventional QCD.
Instead of the RDM Hamiltonian, we consider the conventional Dirac operator, and its localization properties probe the ground state of QCD at different temperatures. It is known \cite{witten1998anti} that the deconfinement phase transition holographically corresponds to the Hawking-Page transition and the holographic dual involves the $AdS_5$ BH at $T>T_c$. Therefore, if the localization properties of the Dirac operator are charged at $T_c$, it will be a probe similar to that used to identify the BH horizon as in the SQCD case.

 The conventional order parameter for the deconfinement phase transition in QCD is the Polyakov loop; however, it was found numerically that the Dirac operator spectrum feels the transition as well. The eigenfunctions of the one-particle 4D Euclidean Dirac Operator 
 \begin{equation}
     \hat{D}(A)\Psi_{\lambda}(x)= \lambda \Psi_{\lambda}(x)
 \end{equation}
  are delocalized at $T<T_{ctit}$ while there is a mobility edge separating the localized and delocalized modes at $T>T_{crit}$, the soft modes are localized 
  \cite{PhysRevD.75.034503,PhysRevD.86.114515,PhysRevD.92.094513}. There is no analytic derivation of this phenomenon yet, but it has solid ground in numerical simulations. Qualitatively, it is explained as the Anderson transition in the instanton-antiinstanton medium.
 The possible holographic explanation of the emergent mobility edge in terms of the string fragmentation near the BH horizon has been suggested in \cite{gorsky2019metal}

The scenario of partial deconfinement has been discussed in \cite{hanada2019anatomy,hanada2019partial} when in the dual holographic language the matrix describing the BH geometry acquires the block-diagonal form. Presumably, it corresponds to the decay of BH into the set of smaller BH.  This emergent fragmentation of the matrix looks similar to our case where the effective clusterization of the modes emerges in the multifractal phases. Since the localization properties of eigenmodes of the Dirac operator in 4D QCD as well as the localization of eigenfunction of the Cooper pair Hamiltonian both play the role of identifier of the phase, we could conjecture that in the partial deconfinement phase the Dirac operator eigenmodes could exhibit fractality. However, it is not completely clear what is the correct counterpart of the instanton-antiinstanton ensemble in the gravity case, although the ensemble of wormholes is the most natural candidate.

The charged black hole holographically corresponds to the boundary theory with the chemical potential, hence we can focus on the phase diagram of the dense QCD. At a small temperature and a large chemical potential, QCD is in the phase of the color superconductivity, and hence indeed the superconducting regime 
is relevant. Since the Dirac operator can be represented by a matrix \cite{verbaarschot2000random}, it would be interesting to investigate its localization properties focusing on the behavior of the $U(1)$ charge. 

Note that in the dense QCD in the CFL phase with $N_c=3$ we have the appropriate ingredients for the analog of the baryonic states in SQCD that we have focused on above. Indeed, there are the vortex strings \cite{eto2009color} which are similar to the vortex strings found in non-SUSY theory in \cite{gorsky2005non} and the magnetic monopoles localized of the vortex strings \cite{gorsky2011confined, eto2011confined}. Hence, it would be interesting to consider the similar monopole pair states in color-flavor locking phase of QCD.

We have seen that in SQCD the origin of the cyclic RG presumably related to the formation of the horizon is the combination of the effects of the chemical potential and the $\theta$ term. Therefore we could expect that the effects of $\theta$-term and anomalies in the dense QCD discussed in \cite{son2001instanton} are of great importance for the formulation of the gravity dual picture involving a charged BH horizon. Note that in the holographic picture for the CFL phase of dense QCD \cite{chen2010towards} the flavor branes touch the horizon, hence
the degrees of freedom populated the flavor brane can be involved in the formation of the horizon.
Remark also that some analog of the cyclic RG is present in dense QCD in a bit different form \cite{son1999superconductivity}.

\subsection{More directions}

\begin{itemize}
    \item It would be interesting to perform a similar analysis for the generic multipair sector of the deterministic and disordered RDM model. In the deterministic model it is possible to consider the different integers $n_i$ in the BA framework and investigate
the aspects of fractality. The total global charge Q governs the tower of Efimov-like states and generically depends on the disorder strength. 
It is desirable to include other parameters in the analysis : $\varepsilon_i$ or the real part of the twist $\chi$ and  solve the BA equations to obtain
$Q(\gamma,\theta,\varepsilon_i,\chi)$. In the SQCD context it would mean the investigation of dependence of  on the quark masses and $\chi =\frac{1}{g_{YM}^2}$.
This could be a small fragment of the general problem of evaluating the marginal stability surfaces for the
BPS states in the $\Omega$-deformed SQCD. It seems that a proper generalization of the cyclic RG
involving more parameters can also be developed along this line of reasoning. We shall discuss these issues in the separate study.

\item Our study shows that the search for fortuitous microstates can be engineered simply by solving the BA equation generalizing the
cohomological arguments behind the R-concentration phenomenon \cite{chang2024fortuity}. There are many examples 
of relations of BA equations with SYM theories with different matter content, hence we expect that similar analysis of BA equations can be 
performed for BPS sectors in such theories as well. It can also be formulated in purely mathematical terms 
as a specific stability structure in the equivariant quantum cohomologies for different manifolds.

    \item We have discussed BPS fractality using the BA approach in 
a specific sector of ${\cal N}=2$ SQCD. It would be interesting to investigate the fractality properties of other BPS sectors or the
    analogous soliton subsectors in the Hilbert spaces of the theories with less amount of SUSY. In particular, it would be interesting to analyze from this viewpoint the instead-of-confinement mechanism \cite{shifman2014lessons} in 
${\cal N}=1$ SQCD when the monopole states are transformed into the quark states.

    \item In our study, the $\theta$ term plays a crucial role. In particular, it fixes the period of the cyclic RG $T^{-1}\propto \sin (\theta_{4d}-\pi)$. In 4d theory $\theta_{4d}=\pi$  is a special point \cite{gaiotto2017theta,gaiotto2018time}, and it would be interesting to apply the formalism of generalized symmetries \cite{gaiotto2015generalized} to this class of problems.

    \item Using the identity of the BA equations, we can immediately conjecture the existence of  the family of Efimov-like scales at the strong coupling point in the large $N_c$ limit of SQCD. For simplicity, assume that the masses are equidistant $m_{i+1}-m_i= \delta$. Then using the solution of the cyclic RG equation for RDM we can write for 4D SQCD the following tower of states.
 \begin{equation}
     \Delta_Q\propto \Delta_0 \exp(- \frac{ \pi Q \delta}{r \sin (\theta_{4D}-\pi)}), \qquad \Delta_0 \propto \exp(-\frac{\delta}{r})
 \end{equation}
 The higher Efimov states are sensitive to the four-dimensional $\theta$ parameter and therefore involve the non-perturbative 
 instanton contributions. Since the $\theta_{4d}=\pi$ point is usually assumed to be the point of spontaneous CP symmetry breaking, we observe that approaching this point in dense matter is nonanalytic. Certainly, the clarification of this point deserves further study.

    \item The phenomenon of fortuity is attributed to some constraints in the algebra of observables at finite N, Usually these constraints are formulated for SU(N) related cohomologies and apply for the identification of BH microstates. In our study, we have a kind of similar relations in terms of quantum equivariant cohomologies of $T^*CP(N-1)$ formulated using BA equations. It is known that similar relations formulated via BAE for twisted inhomogeneous XXZ and XYZ chains can be interpreted in terms of K-theory and elliptic cohomologies, respectively; see, for instance, \cite{koroteev2021quantum,frenkel2023q} and references therein. It would be interesting to develop a similar analysis of fractality in the corresponding cases. We can expect that there are some stability domains in the parameter manifolds. Having in mind the Matsuo-Cherednik duality and the relation of the torus knot invariants with the graded multiplicities of the Calogero-Moser spectrum
    we could expect the relation of fortuity with the 
    stability conditions for the torus knots \cite{stovsic2007homological,gorsky2013stable,gorsky2013quadruply, chauhan2025full}

    \item We have used intensively the fractal dimension $D_q$ to justify the  fractality of the modes in some small 
    subsector of the BPS states. It would be interesting to assign the fractal dimension to generic  BPS networks \cite{gaiotto2013spectral,gaiotto2014spectral,longhi2018wall,galakhov2015spectral}, considering the hopping problem on the corresponding graph. On the other hand, it can be expected that the fractal dimension or the fractality phenomenon in general can be described in the framework of Liouville or Toda field theory along the lines of \cite{kogan1996liouville}.

\end{itemize}

\section{Conclusion}

In this paper, we first determine the phase structure of RDM with and without disorder and find extensive
fractality domains.
Then using the exact equivalence of BA equations apply our findings to formulate
the conjecture of BPS fractality in the
specific subsector of the Hilbert space of strongly coupled ${\cal N}=2$ SQCD with $N_F=2N_C$.

Our main findings are as follows;

\begin{itemize}
    \item Our study provides an interesting mechanism for emergent fractality in deterministic integrable systems based on the exact solution to  BAE. In contrast to the more conventional fractality domain in the disordered system, which can be detected analyzing the statistics of the energy levels in our mechanism, the properties of the other conserved charge play a key role.
    
          This could be a generic situation for integrable systems when chaoticity and fractality cannot be expected naively. Since the RDM Hamiltonian is 
    one of the commuting Hamiltonians of the inhomogeneous spin chains, it suggests that generically the fractality of the sector of the integrable model can be detected by investigating the eigenvectors of selected integral of motion.
    This argument seems to be important in deriving the ETH for the integrable models with BA. Generally speaking, disorder becomes replaced by incommensurativity.

\item In the disordered case the global charge defined by the BAE loses the staircase structure in the multifractal regime but still serves as the proper order parameter. This can be in particular recognized by evaluation of the quantum metrics.

\item  The $\theta$-term somewhat surprisingly plays an important role 
in the stochastic properties of the whole system or in some subsector.
In particular, we provided an example when the transition $\theta\rightarrow \pi$
is non-analytic and $\sin \theta$ defines the inverse period of the cyclic RG flow at large N. Moreover, the multiple non-perturbative scales in the theory are non-analytic in $\theta_{4d}$
and are proportional to $\exp({-\frac{c}{(\theta_{4d}-\pi)}})$.
The $\sin \theta$ also is identified as the coupling constant in the integrable RS model, which hopefully allows us to use the intuition of its multiple reincarnations as an effective theory for topological degrees of freedom, for example, for FQHE.

    \item We conjecture the BPS fractality phenomenon, which corresponds to the intermediate behavior between the BPS chaos and BPS localization for the properly chosen operator in some protected finite sub-sector of the Hilbert space. Our example is very restricted and is based only on the exact solution BA equation
    for the extended defect , but we believe that the phenomenon is quite generic and a similar analysis can be performed for generic BPS networks \cite{gaiotto2013spectral, gaiotto2014spectral, galakhov2015spectral}. In more formal terms, we conjecture that BPS fractality corresponds to the domain in the parameter space with the multiple wall-crossings in the subsector of the Hilbert space. Since BPS states correspond to equivariant quantum cohomologies of the particular manifolds, we expect that mathematically the fractal behavior corresponds to peculiar wall-crossing phenomena for these manifolds. These findings could be useful for the discussion of relevant microstates for the horizon formation in the theory dual to thermal SQCD.

\end{itemize}

The authors thank I.Burmistrov, A. Gerasimov, I. Khaymovich, N. Nekrasov and A. Yung for useful comments. A.G. thanks IHES, where the paper has been completed, for the hospitality and support.

\section{Appendix}
\subsection{Exact solution to the hopping problem}
First, we will derive the expression for the eigenstates as functions of the diagonal elements $\varepsilon_i$ and the parameters $r, \theta$, and obtain one-pair BAE from spectral problem. 
\begin{equation}\label{spectral}
    (H\psi)_i=\sum_{j} H_{ij}\psi_j = \sum_{j<i}(-r)e^{i\theta} \psi_j + \sum_{j>i}(-r)e^{-i\theta}\psi_j + (\varepsilon_i -x)\psi_i = E\psi_i
\end{equation}
Subtracting equation $i$-th from equation $i+1$-th:
\begin{equation}
    re^{i\theta}\psi_{i} - re^{-i\theta}\psi_{i+1} + \varepsilon_{i+1}\psi_{i+1} - (\varepsilon_i - x)\psi_i  = E(\psi_{i+1} - \psi_i)
\end{equation}
we obtain a recurrence relation for $\psi_i$:
\begin{equation}\label{rec}
    \psi_{i+1} = \psi_i \frac{E-\varepsilon_i - iy}{E - \varepsilon_{i+1} +iy }= \psi_i\frac{\rho_{i}}{\rho_{i+1}}e^{-i(\varphi_i+ \varphi_{i+1})}
\end{equation}
with two parameters $\rho_i = \sqrt{(E -\varepsilon_i)^2 + y^2 }$, $\varphi_i = \arctan{\frac{y}{E- \varepsilon_i}}$. Using $\psi_j$ in terms of the previous one, $\psi_{j-1}$ we obtain the expression
\begin{equation}
    \abs{\psi_i} = \frac{\rho_1}{\rho_i}\abs{\psi_1}
\end{equation}
Adding normalization, we derive the following.
\begin{equation}
    \abs{\psi_i} = \frac{1}{\rho_i \sqrt{\sum_{k} \frac{1}{\rho^2_k}}} = 
    \frac{1}{\sqrt{\sum_{k} \frac{1}{\rho^2_k}}}\frac{1}{\sqrt{(E -\varepsilon_i)^2 + y^2 }}
\end{equation}
 We can now derive the BAE considering the first and last equations of the system (\ref{spectral}):
\begin{equation}
    \begin{cases}
        \sum_{i>1} \psi_i (-r e^{i\theta})+(\varepsilon_1 - x) \psi_1 = E\psi_1 \\
        \sum_{i<N} \psi_i (-r e^{-i\theta})+(\varepsilon_N-x) \psi_N = E\psi_N
    \end{cases}
\end{equation}
\begin{equation}
    \begin{cases}
        \psi_N + \sum_{1<i<N} \psi_i +\frac{(\varepsilon_1-x) \psi_1}{(-r e^{i\theta})} = \frac{E\psi_1}{(-r e^{i\theta})} \\
        \psi_1+\sum_{1<i<N} \psi_i +\frac{(\varepsilon_N-x) \psi_N}{(-r e^{-i\theta})} = \frac{E\psi_N}{(-r e^{-i\theta})}
    \end{cases}
\end{equation}
Taking the difference of equations, the following relation arises:
\begin{equation}
    \psi_N = \psi_1 e^{-2i\theta}\frac{E - \varepsilon_1 - iy}{E - \varepsilon_N +iy}
\end{equation}
Using equation (\ref{rec}), we can express $\psi_N$ in terms of $\psi_1$ to get 
\begin{equation}
    e^{-2i\theta}\prod_{l = 1}^{N} \frac{E - \varepsilon_l - iy}{E - \varepsilon_l + iy} = 1
\end{equation}
For $\varepsilon_i = 0$, the spectrum is indexed by the integer $0\leq k\leq N-1$:
\begin{equation}\label{E_free}
    E_k  =N^{-\gamma}\sin{\theta}\cot \frac{\pi k-\theta}{N} 
\end{equation}
 and (\ref{rec}) reduces to 
\begin{equation}
    \psi_{i+1} = \psi_i \frac{E_k - iy}{E_k +iy} = \psi_i e^{-2i\phi_k}
\end{equation}
where $\phi_Q = \arctan{(r\sin{\theta}/E_k )} = \frac{\pi Q - \theta}{N}$. The eigenstates have the form of plane waves with momentum $p = 2\phi_Q$
\begin{equation}\label{eq:plain_waves}
\ket{Q} = \sum_n \frac{e^{-2i\phi_Q n}}{\sqrt{N}} \ket{n} 
\end{equation}
\subsection{Evaluation of the fractal dimension}
Now we have sum $I_q$ for the eigenstate in the following form:
\begin{equation}\label{Iq_exact}
    I_q = \sum_{i} \frac{1}{(\rho_i \sqrt{C})^{2q}}
\end{equation}
where $C = \sum_{i}\frac{1}{\rho^2_i}$. We take $\delta = \omega/N$ as in \cite{anfossi2005elementary}, where the diagonal elements $\varepsilon_n/2$ were within the Debye shell of width $2\omega_c$. Consequently, the diagonal elements are given by $\varepsilon_i = \frac{\omega}{N}(i - \frac{N}{2})$. The probability distribution profile for the eigenstate with energy $E$ has a scale of $\Gamma$ with a characteristic number of sites $n_{s} = {N\Gamma}/\omega$. To replace the sum with a integral in (\ref{Iq_exact}) the partition mesh must be smaller than the characteristic scale of the function, which requires a large number of sites in $n_s$: ${N\Gamma}/\omega \gg 1$. This condition implies $\gamma < 1$, we will see that $\gamma = 1$ corresponds exactly to the point of localized-fractal transition.
\[
    C = \sum_{i}\frac{1}{(E -\varepsilon_i)^2 + y^2} \to \frac{N}{\omega}\int_{-\omega/2}^{\omega/2}d\xi \frac{1}{(E -\xi)^2 + y^2} = 
\]

\begin{equation}
    =\frac{N}{\omega}\frac{1}{y}(\arctan{\frac{-E + \omega/2}{y}} -\arctan{\frac{-E - \omega/2}{y}})
\end{equation}
\[
    I_q = \frac{1}{C^q}\sum_{i} \frac{1}{((E -\varepsilon_i)^2 + y^2)^q} \to \frac{N}{\omega C^q}\int_{-\omega/2}^{\omega/2}d\xi \frac{1}{((E -\xi)^2 + y^2)^{q}} = 
\]
\[
    = \frac{N}{\omega C^q}(\frac{-E + \omega/2}{y^{2q}} {}_2 F_1 (\frac{1}{2}, q,\frac{3}{2}, -\frac{(-E + \omega/2)^2}{y^2}) - 
\]
\begin{equation}
    - \frac{-E - \omega/2}{y^{2q}} {}_2 F_1 (\frac{1}{2}, q,\frac{3}{2}, -\frac{(-E - \omega/2)^2}{y^2}))
\end{equation}
To obtain the fractal dimensions analytically, we will find the asymptotic form of the integral:
\begin{equation}
    \frac{N}{\omega}\int_{-\omega/2}^{\omega/2}d\xi \frac{1}{((E -\xi )^2 + y^2)^q} 
\end{equation}
Changing variables and taking large $N$ for $\gamma \in (0,1)$: 
\begin{equation}
    \frac{N}{\omega} y^{1-2q}\int_{-(\omega/2 - x +E )/y}^{(\omega/2+x-E)/y}d\xi \frac{1}{(\xi^2 + 1)^q} \to \frac{N}{\omega} y^{1-2q}\int_{-\infty}^{\infty}d\xi \frac{1}{(\xi^2 + 1)^q} =  \frac{N}{\omega} y^{1-2q}\sqrt{\pi}\frac{\Gamma(q - 1/2)}{\Gamma(q)}
\end{equation}
The validity of the approximation can be checked by estimating the tail of the integral, $\int_{N^{\gamma}}^{\infty} 1/\xi^{2q}\sim N^{\gamma (1-2q)}$. This term is parametrically smaller than the full integral $\int^{\infty}_{-\infty}$ for $q > 1/2$, which is also an essential condition for the integral's convergence.
\begin{equation}
    I_q = \frac{N^{1-(1-2q)\gamma} \sin^{1-2q}{\theta}\sqrt{\pi}\frac{\Gamma(q-1/2)}{\Gamma(q)}}{(N^{1+\gamma}\pi \sin{\theta})^q} = N^{(1-q)(1-\gamma)}\pi^{\frac{1}{2}(1-2q)}\frac{\Gamma(q - 1/2)}{\Gamma(q)}\sin^{1-q}{\theta}
\end{equation}
Note that integral tails produce corrections of the order $1/(N^{\alpha}\ln{N})$ to expression (\ref{D_q_clean}), which decrease much faster than $1/\ln{N}$ in the limit of $N \to \infty$.
\subsection{Metric for $N = 2$ case }

Consider the simplest case of $N=2$, which is similar to the discussion in \cite{sharipov2024hilbert} with
a few minor differences. We use a uniform, shifted distribution for diagonal elements. From (\ref{scalar}), we see that it is sufficient to shift only the first element: $\varepsilon_1 \in (-W/2+\delta, W/2+\delta)$, $\varepsilon_2 \in (-W/2, W/2)$, contrary to
the normal distribution used in \cite{sharipov2024hilbert}, so the measure is $d\mu = \frac{1}{W^2}d \varepsilon_1 d \varepsilon_2$. 
The quantum metric for $N = 2$, along with the eigenstates and energies, can be calculated explicitly. We expect that for a large shift $\delta$, the denominator in the metric definition will be regularized by $\delta$. Therefore, we can monitor the behavior of the metric near $r = 0$:
\[
G_{rr} = \frac{1}{4r W^2}(2 r \ln{\frac{(\delta^2+4 r^2)^2}{((\delta-W)^2+4 r^2)((\delta+W)^2+4 r^2)}} +(\delta-W) \arccot{\frac{2 r}{\delta-W}}+
\]
\begin{equation}
    +(\delta+W) \arccot{\frac{2 r}{\delta+W}}-2 \delta \arccot{\frac{2 r}{\delta})}
\end{equation}
With expansion for $\delta < W$:
\[
\frac{\pi}{4r W^2}(W - \delta) + O(r)
\]
and for $\delta > W$: 
\[
\frac{1}{2W^2} \ln{\frac{\delta^4}{(\delta - W)^2(\delta + W)^2}} + O(r^2)
\]
Analogously:
\[
G_{\theta\theta} = \frac{r}{2 W^2}(r \ln{\frac{(\delta^2+4 r^2)^2}{((\delta-W)^2+4 r^2)((\delta+W)^2+4 r^2)}} +(\delta-W) \arccot{\frac{2 r}{\delta-W}}+
\]
\begin{equation}
    +(\delta+W) \arccot{\frac{2 r}{\delta+W}}-2 \delta \arccot{\frac{2 r}{\delta})}
\end{equation}
The transition occurs when the support of the distribution stops intersecting the line of degeneracy ($\varepsilon_1 = \varepsilon_2$) for the unperturbed model, and the singular behavior for $G_{rr}$ and $G_{\theta\theta}$ becomes constant. There is also a transition in the embedding: near the point $r = 0$ with the condition: $\delta > W $:
\begin{equation}
    G_{rr}\sim const + r^2,\;\;\; G_{\theta\theta} \sim r^2(const + r^2)
\end{equation}
 It corresponds to a finite curvature at $r = 0$, while for $\delta < W$ the embedding is a cone with a curvature singularity at $r = 0$.
\begin{figure}[h]
\begin{minipage}[h]{0.49\linewidth}
\center{\includegraphics[width=1\linewidth]{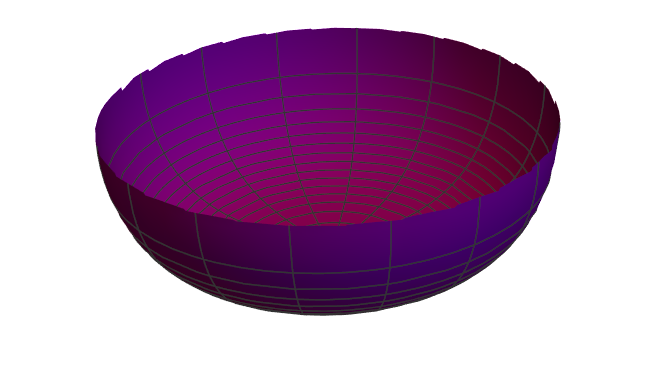} \\ $\delta > W$}
\end{minipage}
\hfill
\begin{minipage}[h]{0.49\linewidth}
\center{\includegraphics[width=1\linewidth]{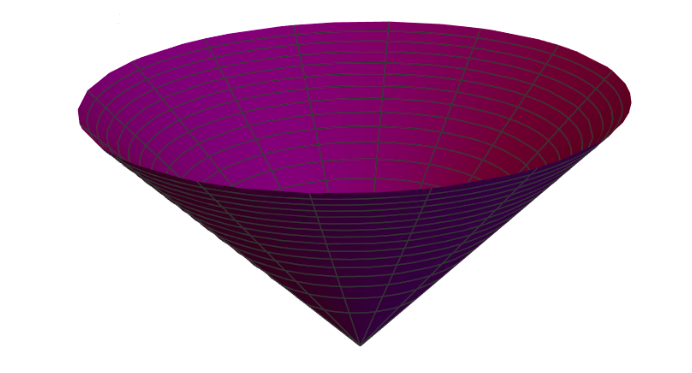} \\$\delta < W$}
\end{minipage}
\caption{Embeddings of isometric manifold for different $\delta$}
\label{ris:image2}
\end{figure}

\subsection{Berry curvature, $N=2$}

Let us now turn to the evaluation of the Berry curvature for the case $N=2$.
From the definition of quantum metric $g_{\alpha \beta}$ the only component that can have an imaginary part is $g_{xy}$. Integration over diagonal elements distribution yields the following expression for the Berry curvature:
\begin{equation}
    \Omega_{xy} = \frac{1}{2W^2}\ln{\frac{(\sqrt{4r^2+(\delta - W)^2}+W - \delta)(\sqrt{4r^2+(\delta + W)^2} - W - \delta)}{(\sqrt{\delta^2+4r^2} - \delta)^2 }}
\end{equation}
In case of identical distributions of $\varepsilon_1$ and $\varepsilon_2$, $\delta = 0$, the Berry curvature vanishes. The integral of the Berry curvature on $x$ and $y$ gives the following:
\begin{equation}
    I(\delta, W)=\int \Omega_{xy} dx dy = 
    \begin{cases}
    \frac{\pi\delta}{2W^2} (2W - \delta) \;\;\;\;\; \delta < W   \\
    \frac{\pi}{2}    \;\;\;\;\; \delta > W
    \end{cases}
\end{equation}
\begin{figure}[H]
    \centering
    \includegraphics[width=0.6\linewidth]{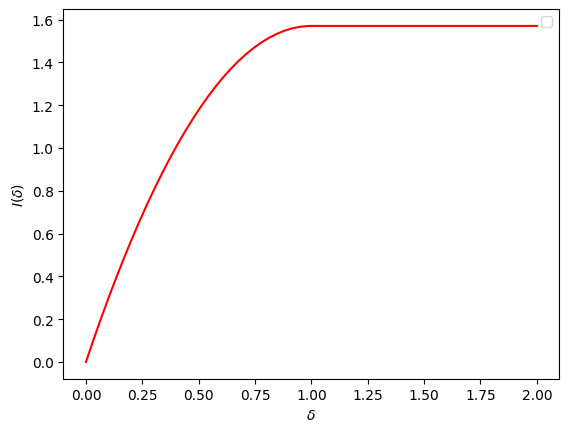} 
    \caption{$ \int \Omega_{xy} (\delta) dx dy$ for $W = 1$} 
    \label{fig:my_label3}
\end{figure}

\subsection{Metric at small $r$ at $N=2$ }
To evaluate a metric for the $N = 2$ case, we introduce $\zeta = \varepsilon_1 - \varepsilon_2$, $\varepsilon_2$, then the metric can be found as:
 \begin{equation}
     G_{rr} = \int (1 - \frac{\abs{\zeta - \delta}}{W})\frac{d\zeta}{W} \frac{(\frac{1}{4}(\zeta - \sqrt{\zeta^2 + 4r^2})^2 - r^2)^2}{(2r^2 + \frac{1}{2}\zeta^2 - \zeta\sqrt{\zeta^2 + 4r^2})^2(\zeta^2 + 4r^2)}  
 \end{equation}
The expression for the metric reads as 
\[
G_{rr} = \frac{1}{4r W^2}(2 r \ln{\frac{(\delta^2+4 r^2)^2}{((\delta-W)^2+4 r^2)((\delta+W)^2+4 r^2)}} +(\delta-W) \arccot{\frac{2 r}{\delta-W}}+
\]
\begin{equation}
    +(\delta+W) \arccot{\frac{2 r}{\delta+W}}-2 \delta \arccot{\frac{2 r}{\delta})}
\end{equation}
which has the following expansion for $\delta < W$:
\[
\frac{\pi}{4r W^2}(W - \delta) + O(r)
\]
and for $\delta > W$: 
\[
\frac{1}{2W^2} \ln{\frac{\delta^4}{(\delta - W)^2(\delta + W)^2}} + O(r^2)
\]
Analogously for the component of the metric $G_{\theta\theta}$:
 \begin{equation}
     G_{\theta\theta} = \int (1 - \frac{\abs{\zeta - \delta}}{W})\frac{d\zeta}{W} \frac{(\frac{1}{4}(\zeta - \sqrt{\zeta^2 + 4r^2})^2 + r^2)^2}{(2r^2 + \frac{1}{2}\zeta^2 - \zeta\sqrt{\zeta^2 + 4r^2})^2(\zeta^2 + 4r^2)}  
 \end{equation}
\[
G_{\theta\theta} = \frac{r}{2 W^2}(r \ln{\frac{(\delta^2+4 r^2)^2}{((\delta-W)^2+4 r^2)((\delta+W)^2+4 r^2)}} +(\delta-W) \arccot{\frac{2 r}{\delta-W}}+
\]
\begin{equation}
    +(\delta+W) \arccot{\frac{2 r}{\delta+W}}-2 \delta \arccot{\frac{2 r}{\delta})}
\end{equation}
The small $r$ behavior of the metric changes when the support of the distribution does not intersect the line of degeneracy ($\varepsilon_1 = \varepsilon_2$) (for an unperturbed model), and in this case, the singular behavior for $G_{rr}$ and $G_{\theta\theta}$ disappears. There is a similar transition in embedding: if $\delta > W $ 
\begin{equation}
    G_{rr}\sim const + r^2,\;\;\; G_{\theta\theta} \sim r^2(const + r^2)
\end{equation}
  The curvature is finite in $r = 0$. For $\delta < W$, embedding can be obtained via restriction of the Euclidean metric on a cone:
\begin{equation}\label{eq:7}
    Z^2 - \frac{1- \alpha^2}{\alpha^2} (x^2 + y^2) = 0
\end{equation}
\begin{equation}
    R^2(r) = 
\frac{\pi r}{2 W^2}(W - \delta) + O(r^3)
\end{equation}
\begin{equation}
    \left(\frac{dZ}{dr}\right)^2 = G_{rr} - \left(\frac{dR}{dr}\right)^2 = \frac{\pi }{4r W^2}(W - \delta) - \frac{\pi }{8r W^2}(W - \delta) = \frac{\pi}{8r W^2}(W - \delta)
\end{equation}
Solving these equations, in the vicinity of $r = 0$:
\begin{equation}
    Z = \sqrt{\frac{\pi r}{2 W^2}(W - \delta)}\;\;\;\;\;R = \sqrt{\frac{\pi r}{2 W^2}(W - \delta)}
\end{equation}
near $r = 0$ $\frac{dZ}{dR} = 1$, and we have a conical singularity.
Analogously to the metric, the Berry curvature can be found as
\begin{equation}
    \Omega_{xy} = \int \frac{d \lambda_1 d \lambda_2}{W^2} \frac{\frac{1}{8}(\zeta - \sqrt{\zeta^2 + 4r^2})^4 - 2r^4}{(\zeta^2 + 4r^2)(4r^2 + \zeta^2 - \zeta\sqrt{\zeta^2 + 4r^2})^2}
\end{equation}
 where $\zeta = \varepsilon_1 - \varepsilon_2$.

\subsection{Metric at small $r$ in the large $N$ limit}

Let us evaluate the metric near $r=0$ at large $N$
considering the diagonal part as an unperturbed Hamiltonian and the off-diagonal part as a perturbation.
We will substitute the sum over $n$ by an integral :
\begin{equation}
    \sum_n \to \int_0^N
\end{equation}
For $G_{rr}$, one obtains:
\begin{equation}
G_{rr} = \sum_{n = 1}^N \sum_{m\neq n} \frac{1}{(n-m)^2\delta^2}= \frac{2}{\delta^2 N}(\frac{\pi^2}{6}N - \ln{N}+O(1/N))
\end{equation}

and
\begin{equation}
    G_{\theta\theta} = r^2  G_{rr}
\end{equation}
However, to evaluate $G_{r\theta}$, we need to consider higher orders of $z = {r e^{i\theta}}$. The first correction for the eigenstates $\ket{\psi_n^{1}} = \sum_{m\neq n} c_m\ket{m}$ is as follows:
\begin{equation}
    c_{m\neq n}^{(1)} = \begin{cases}
        -\frac{\bar{z}}{(n-m)\delta} \;\;\; m>n\\
        -\frac{z}{(n-m)\delta}\;\;\;m<n
    \end{cases}
\end{equation}
and the first non-zero contribution for $G_{r\theta}$ is
\begin{equation}
    \bra{\psi_n^{(1)}}\partial_r H\ket{\psi_m^{(1)}} \bra{\psi_m^{(1)}}\partial_\theta H\ket{\psi_n^{(1)}} 
\end{equation}
Firstly, let us compute the multiplier with $\partial_r H$:
\begin{equation}
    \bra{\psi_m^{(1)}}\partial_r H\ket{\psi_n^{(1)}} = \bra{b}\ket{c_n^{(1)}}
\end{equation}
and obtain $\bra{b}$ for $m>n$:
\begin{equation}
    b_k = \begin{cases}
        k < n \;\;\;\frac{\abs{z}^2}{\delta}\sum_{l=1}^{k-1} \frac{1}{n - l} + 
        \frac{\bar{z}^2}{\delta}\sum_{l=1}^{n-k-1} \frac{1}{l} - \frac{\abs{z}^2}{\delta}\sum_{l=1}^{N-n} \frac{1}{l}\\
         k = n\;\;\;\frac{\abs{z}^2}{\delta}\sum_{l=1}^{n-1} \frac{1}{l} - \frac{\abs{z}^2}{\delta}\sum_{l=1}^{N-n} \frac{1}{l}\\
        k > n\;\;\;\frac{\abs{z}^2}{\delta}\sum_{l=1}^{n-1} \frac{1}{n - l} - 
        \frac{z^2}{\delta}\sum_{l=1}^{k-n-1} \frac{1}{l} - \frac{\abs{z}^2}{\delta}\sum_{l=k+1}^{N-n} \frac{1}{l}
    \end{cases}
\end{equation}
Analogously, we can calculate:
\begin{equation}
    \bra{\psi_m^{(1)}}\partial_\theta H\ket{\psi_n^{(1)}} = \bra{d}\ket{\psi_n^{(1)}}
\end{equation}
With components of $\bra{d}$ for $m>n$:
\begin{equation}
    d_k = \begin{cases}
        k < n \;\;\;i\frac{\abs{z}^2}{\delta}\sum_{l=1}^{k-1} \frac{1}{n - l} -
        i\frac{\bar{z}^2}{\delta}\sum_{l=1}^{n-k-1} \frac{1}{l} + i\frac{\abs{z}^2}{\delta}\sum_{l=1}^{N-n} \frac{1}{l}\\
         k = n\;\;\;i\frac{\abs{z}^2}{\delta}\sum_{l=1}^{n-1} \frac{1}{l} + i\frac{\abs{z}^2}{\delta}\sum_{l=1}^{N-n} \frac{1}{l}\\
        k > n\;\;\;i\frac{\abs{z}^2}{\delta}\sum_{l=1}^{n-1} \frac{1}{n - l} - 
        i\frac{z^2}{\delta}\sum_{l=1}^{k-n-1} \frac{1}{l} + i\frac{\abs{z}^2}{\delta}\sum_{l=k+1}^{N-n} \frac{1}{l}
    \end{cases}
\end{equation}
Since $N\gg 1$, we replace the sum with integral (assuming $m$ and $n$ $\gg 1$) and find:
\begin{equation}
    G_{r\theta} = A r^5 \sin{2\theta} = \widetilde{A}\frac{\ln^4{N}}{\delta^4}r^5 \sin{2\theta}
\end{equation}
- where $\widetilde{A} \sim O(1)$ and independent of $N$ in leading order. Taking into account that $G_{rr} = C + \alpha r^2$, $G_{\theta\theta} = Cr^2 - \beta r^4$, $G_{r\theta} = -A r^5 \sin{2\theta}$ we derive the Gaussian curvature.
\begin{equation}
    K = \alpha - \beta + \frac{4\beta}{C^2}
\end{equation}

\subsection{Transfer matrix of twisted inhomogeneous XXX spin chain and RDM Hamiltonian}

The BAE for the RDM model and in inhomogeneous twisted XXX spin chains are identical, 
so it is not surprising that the RDM Hamiltonian can be obtained from the 
spin chain transfer matrix \cite{bork2015particle}. Consider the R-matrix 
\begin{equation}
    R(\lambda)=\frac{1}{\lambda+iy}(\lambda 1\otimes1 +iyP)
\end{equation}
acting in the product of two linear spaces and $P$ is a permutation operator.
It obeys the Yang-Baxter equation,
\begin{equation}
    R_{12}(\lambda_1 -\lambda_2)R_{13}(\lambda_1)R_{23}(\lambda_2)=
    R_{23}(\lambda_2)R_{13}(\lambda_1)R_{12}(\lambda_1 -\lambda_2)
\end{equation}
and in terms of the Cooper pair pseudospin operators, it can be presented in the form
\begin{equation}
R = \frac{1}{\lambda +iy} \begin{pmatrix}
\lambda1 +iy\hat{N} & iyb \\
iyb^{\dagger} & \lambda 1 + iy(-\hat{N} +1)
\end{pmatrix}     
    \end{equation}

The monodromy matrix is defined as 
\begin{equation}
    T(\lambda)= \omega_0 R_{0N}(\lambda-\varepsilon_N) \dots R_{01}
    (\lambda-\varepsilon_1)
\end{equation}
where $\omega_0=\exp(i\theta \sigma_3)$, $R_{0i}$ acts on $V_0\otimes V_i$,
both $V_0,V_i$ - auxiliary and physical spaces are $C^2$. 
It obeys the standard RTT relation in the integrable systems
\begin{equation}
    R_{12}9u-v)T(_1(u)T_2(v)=T_2(v)T_1(u)R_{12}(u-v)
\end{equation}

Taking trace over the auxiliary space $V_0$ we get the transfer matrix
\begin{equation}
    t(\lambda)=Tr_0 T(\lambda)
\end{equation}
acting on $V_N=\otimes V_i$ and obeying the relation; 
\begin{equation}
    [t(\lambda_1),t(\lambda_2)]=0, u,v \in C
\end{equation}
After the simple rescaling  rescaling 
\begin{equation}
t(u) \rightarrow \prod_{i=1} ^{N} \frac{u-\varepsilon_i +iy}{u-\varepsilon_i +iy/2} t(u)
\end{equation}
$t(\lambda) $ becomes  the  generating function for  commuting Hamiltonians 
\begin{equation}
    t(\lambda)= \sum_{k=1}\lambda^{-k}t_k,\qquad [t_i,t_k]=0
\end{equation}

Using the expansion of the transfer matrix at infinity 
we obtain for the second non-local Hamiltonian
\begin{equation}
   t_2\propto y \sin\theta \sum_{i=1}^{N}\varepsilon_i \hat{N}_i - \frac{y^2}{2}
   \sum_{i<k}^{N}(e^{i\theta}b^{\dagger}_ib_k + e^{-i\theta}b^{\dagger}_kb_i)+\mathrm{const} 
\end{equation}
which is proportional to the RDM Hamiltonian.

This expression is an analog of formula (\ref{richR}) for the 
expression of the Richardson Hamiltonian from the Gaudin
transfer matrix. Note that the non-local Hamiltonians
derived by the expansion of the transfer matrix at infinity
usually form the Yangian or quantum affine algebra.

\bibliographystyle{unsrt}
\bibliography{references}
\end{document}